\documentclass[11pt,a4paper]{article}

\usepackage[utf8]{inputenc}
\usepackage[T1]{fontenc}
\usepackage{lmodern}
\usepackage{amsmath, amssymb, amsfonts, mathrsfs}
\usepackage{graphicx}
\usepackage{geometry}
\usepackage{tcolorbox}
\usepackage{tikz}
\usepackage{pgfplots}
\pgfplotsset{compat=1.18}
\usetikzlibrary{decorations.markings, positioning, arrows.meta, decorations.pathmorphing, shadows}

\usepackage{hyperref} 

\hypersetup{
	colorlinks=true,
	linkcolor=blue,
	filecolor=magenta,      
	urlcolor=cyan,
}

\DeclareMathOperator{\sech}{sech}
\numberwithin{equation}{section}

\title{\textbf{A Primer on Instanton Physics \\ A Short Course}}
\author{Pervez Hoodbhoy \\
	Lectures delivered at the University of New Brunswick, 5-21 October, 2026}
\date{}

\begin{document}
	
	\maketitle
	
	\begin{quote}
		\textit{``Vacuum decay is the ultimate ecological catastrophe; in a new vacuum there are new constants of nature; after vacuum decay, not only is life as we know it impossible, so is chemistry as we know it. However, one could always draw stoic comfort from the possibility that perhaps in the course of time the new vacuum would sustain, if not life as we know it, at least some creatures capable of knowing joy. This possibility has now been eliminated.''} \\
		-- Sidney Coleman, Frank De Luccia (1980) 
	\end{quote}
	
	\noindent For near half a century, Coleman's grim portent of how the universe could suddenly collapse has inspired physicists to explore the mathematics of quantum tunneling, a process most efficiently understood via the theory of instantons. While some have sought to refine calculations for exactly how long the universe will take to tunnel to its real vacuum state, the overwhelming majority of physicists turned their attention to immediately observable phenomena. The double-well potential is the paradigmatic starting point that served to launch seminal developments in instanton calculus during the 1970s. Thereafter non-perturbative semiclassical methods became indispensable for uncovering the topological properties of Yang-Mills gauge theories and the spectral structures of quantum systems. Instanton methods are now widely used in chemical physics and neural networks.
	
	These six lectures are aimed at senior undergraduate and beginning graduate students. Whereas excellent review materials exist, I felt that a more palatable route might be to bypass the rather detailed discussion of reflectionless potentials, concentrating instead on the Gelfand-Yaglom method for calculating functional determinants.   The first four lectures deal solely with particle mechanics, building the necessary background for the last two that cover applications in semi-classical field theory. 
	
	\setcounter{tocdepth}{1}
	\tableofcontents
	\newpage
	
	\part{Introduction to Quantum Tunneling and Path Integrals}
	
	\section{Quantum Tunneling and the WKB Approximation}
\begin{figure}[h]
		\centering
		\hspace*{-0.5cm} 
		\includegraphics[scale=.40]{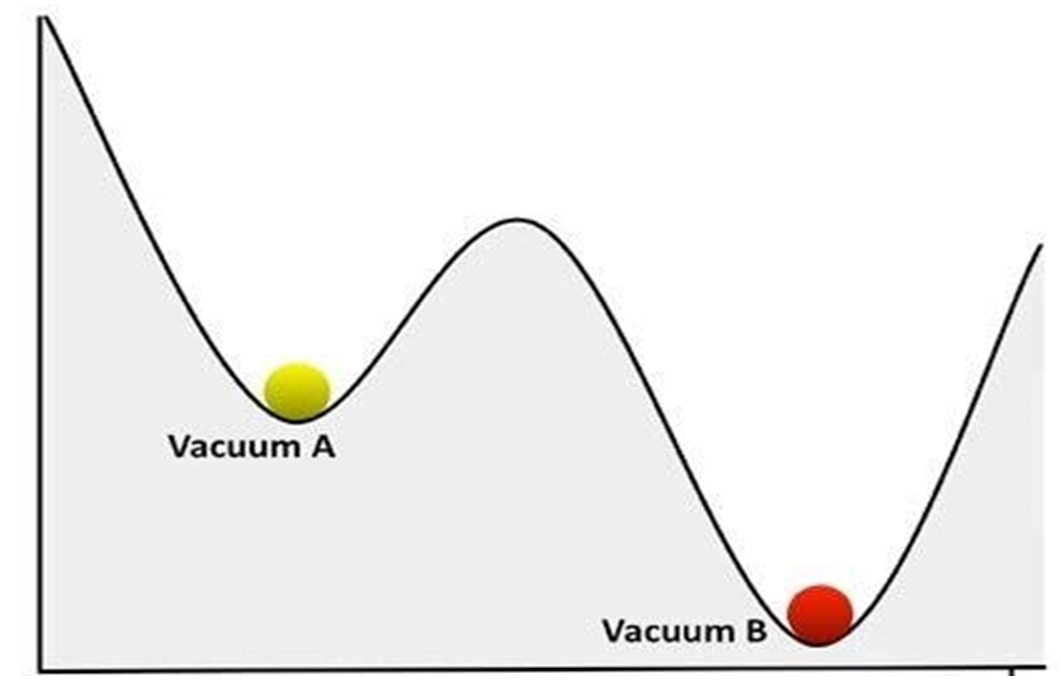}
		\caption{Decay of false vacuum}
		\label{false}
	\end{figure}

	Consider a particle trapped in a local minimum (Vacuum A) separated from a lower energy global minimum (Vacuum B) by a potential energy barrier, Fig.~\ref{false}. Classically, the particle must possess enough kinetic energy to go \textit{over} the potential barrier. Quantum mechanically, the particle can tunnel \textit{through} the barrier, a phenomenon completely forbidden in classical physics.
	
	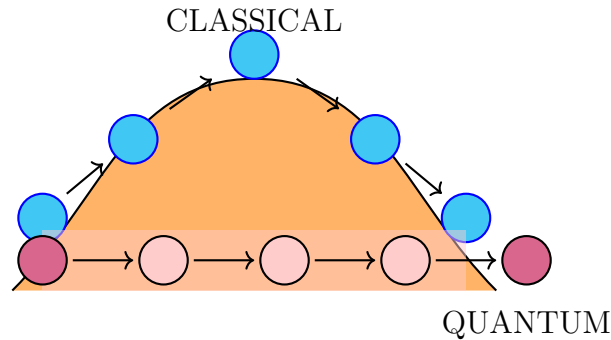
\begin{figure}[h]
		\centering
		\begin{tikzpicture}[scale=0.8]
			\fill[orange!60] (0,0) to[out=45,in=180] (4,3.5) to[out=0,in=135] (8,0) -- cycle;
			\draw[thick] (0,0) to[out=45,in=180] (4,3.5) to[out=0,in=135] (8,0);
			
			\filldraw[cyan!60, draw=blue, thick] (0.5, 1.2) circle (0.4);
			\filldraw[cyan!60, draw=blue, thick] (2, 2.5) circle (0.4);
			\filldraw[cyan!60, draw=blue, thick] (4, 3.9) circle (0.4) node[above=5pt, black] {\large CLASSICAL};
			\filldraw[cyan!60, draw=blue, thick] (6, 2.5) circle (0.4);
			\filldraw[cyan!60, draw=blue, thick] (7.5, 1.2) circle (0.4);
			
			\draw[->, thick] (0.9, 1.6) -- (1.5, 2.1);
			\draw[->, thick] (2.6, 3.0) -- (3.3, 3.5);
			\draw[->, thick] (4.7, 3.5) -- (5.4, 3.0);
			\draw[->, thick] (6.5, 2.1) -- (7.1, 1.6);
			
			\fill[pink!80, opacity=0.5] (0.5, 0) rectangle (7.5, 1);
			
			\filldraw[purple!60, draw=black, thick] (0.5, 0.5) circle (0.4);
			\filldraw[pink!80, draw=black, thick] (2.5, 0.5) circle (0.4);
			\filldraw[pink!80, draw=black, thick] (4.5, 0.5) circle (0.4);
			\filldraw[pink!80, draw=black, thick] (6.5, 0.5) circle (0.4);
			\filldraw[purple!60, draw=black, thick] (8.5, 0.5) circle (0.4) node[below=15pt, black] {\large QUANTUM};
			
			\draw[->, thick] (1, 0.5) -- (2, 0.5);
			\draw[->, thick] (3, 0.5) -- (4, 0.5);
			\draw[->, thick] (5, 0.5) -- (6, 0.5);
			\draw[->, thick] (7, 0.5) -- (8, 0.5);
		\end{tikzpicture}
		\caption{Classical particle climbing the barrier vs. Quantum particle tunneling through the barrier.}
	\end{figure}
	
	\begin{figure}[h]
		\centering
		\includegraphics[width=0.6\textwidth]{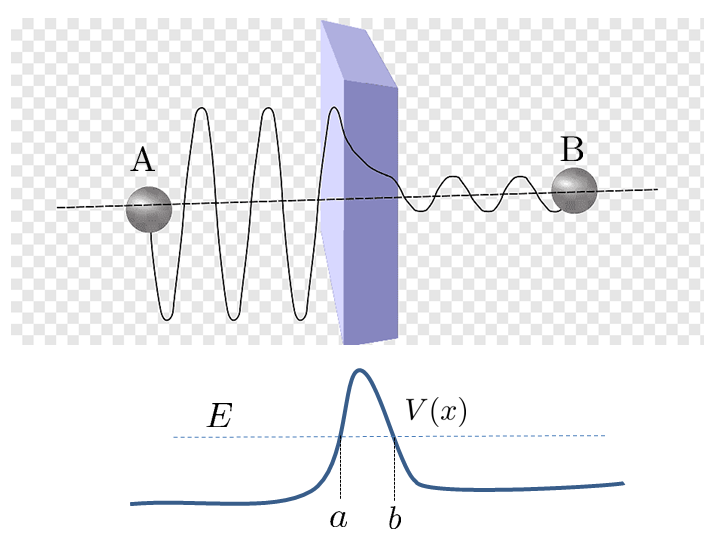}
		\caption{A visual representation of a wave packet tunneling through a finite potential barrier. Note the exponentially decaying amplitude inside the barrier region.}
	\end{figure}
	
	\subsection{WKB Approximation}
	For a particle tunneling through a potential barrier $V(x)$ with energy $E$, the Wentzel-Kramers-Brillouin (WKB) approximation provides an estimate for the tunneling amplitude $\mathcal{A}_{A \rightarrow B}$: 
	\begin{equation}
		\mathcal{A}_{A \rightarrow B} \sim e^{-\frac{2}{\hbar} \int_a^b dx \sqrt{V(x)-E}}
	\end{equation}
	Here, $a$ and $b$ represent the classical turning points where the particle's energy equals the potential energy: $V(a) = V(b) = E$. 
	
	\textbf{Limitations:} The integral evaluates the action over the classically forbidden region. While fundamentally non-perturbative, it is difficult to improve systematically using standard undergraduate quantum mechanics. An alternative to WKB is provided by the Feynman path integral.
	
	\section{The Feynman Path Integral}
	Richard Feynman generalized the concept of the double-slit experiment. 
	
	\begin{figure}[h]
		\centering
		\hspace*{-0.5cm} 
		\includegraphics[scale=.60]{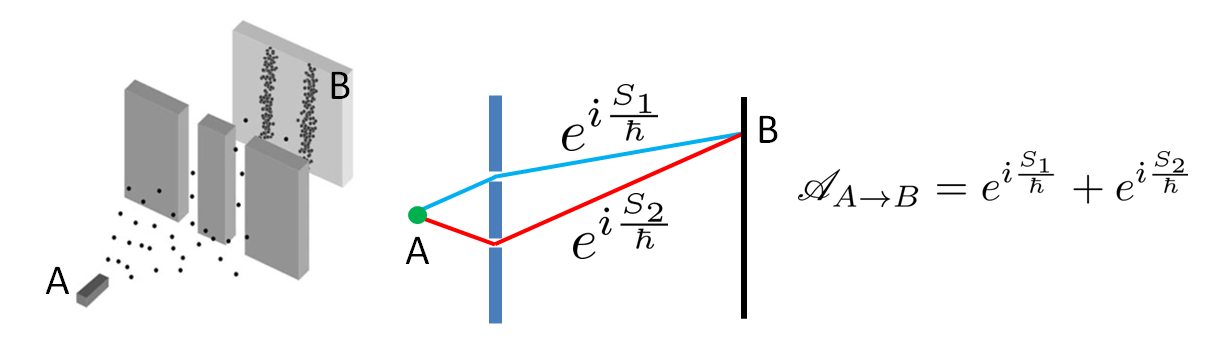}
		\caption{Double Slit Experiment.}	
		\label{double}
	\end{figure}
	
	The quantum amplitude to reach a specific point on the screen is the sum of the amplitudes for the path through slit 1 and the path through slit 2: 
	\begin{equation}
		\mathcal{A}_{A \rightarrow B} = e^{i\frac{S_1}{\hbar}} + e^{i\frac{S_2}{\hbar}}
	\end{equation}
	If we introduce more slits (and more screens), the particle can take many paths: 
	\begin{equation}
		\mathcal{A}_{A \rightarrow B} = e^{i\frac{S_1}{\hbar}} + e^{i\frac{S_2}{\hbar}} + e^{i\frac{S_3}{\hbar}} + e^{i\frac{S_4}{\hbar}} + \dots
	\end{equation}
	
	\subsection{The Continuum Limit: Summing Over All Paths}
	Taking this logic to the extreme, if we have infinitely many screens with infinitely many slits, the particle simultaneously travels on \textit{all possible paths} between the initial and final states. 
	
	\begin{figure}[h]
		\centering
		\begin{tikzpicture}[scale=0.9]
			\draw[thick, ->] (0,0) -- (10,0) node[right] {Space};
			\draw[thick, ->] (0,0) -- (0,5) node[above] {Time};
			\draw[thick] (-0.2, 1) -- (0.2, 1) node[left=5pt] {$t_i$};
			\draw[thick] (-0.2, 4) -- (0.2, 4) node[left=5pt] {$t_f$};
			\draw[thick] (-0.2, 2.5) -- (0.2, 2.5) node[left=5pt] {$t$};
			
			\filldraw (4, 1) circle (0.06) node[below right] {$|x_i\rangle$};
			\filldraw (6, 4) circle (0.06) node[above right] {$|x_f\rangle$};
			
			\draw[thick, dashed] (0, 2.5) -- (10, 2.5);
			
			\foreach \x in {2, 4, 6, 8} {
				\draw[ultra thick] (\x-0.6, 2.5) -- (\x+0.6, 2.5);
			}
			\node at (2, 2.8) {$x_1$};
			\node at (4, 2.8) {$x_2$};
			\node at (6, 2.8) {$x_3$};
			\node at (8, 2.8) {$x_N$};
			
			\draw[->, thick] (4,1) -- (2,2.5); \draw[->, thick] (2,2.5) -- (6,4);
			\draw[->, thick] (4,1) -- (4,2.5); \draw[->, thick] (4,2.5) -- (6,4);
			\draw[->, thick] (4,1) -- (6,2.5); \draw[->, thick] (6,2.5) -- (6,4);
			\draw[->, thick] (4,1) -- (8,2.5); \draw[->, thick] (8,2.5) -- (6,4);
		\end{tikzpicture}
		\caption{Paths through a single intermediate time slice.}
	\end{figure}
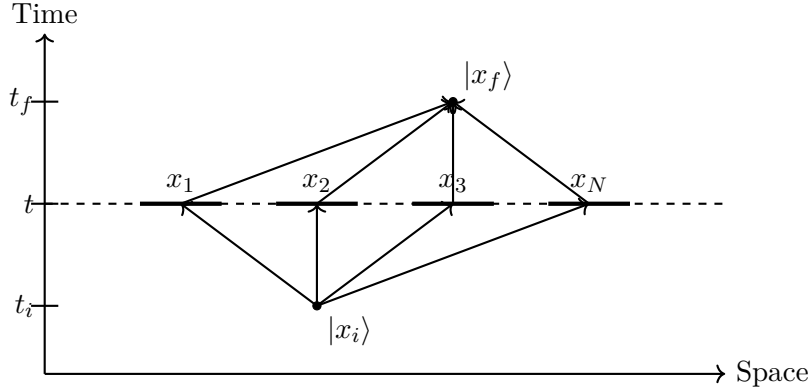
	
	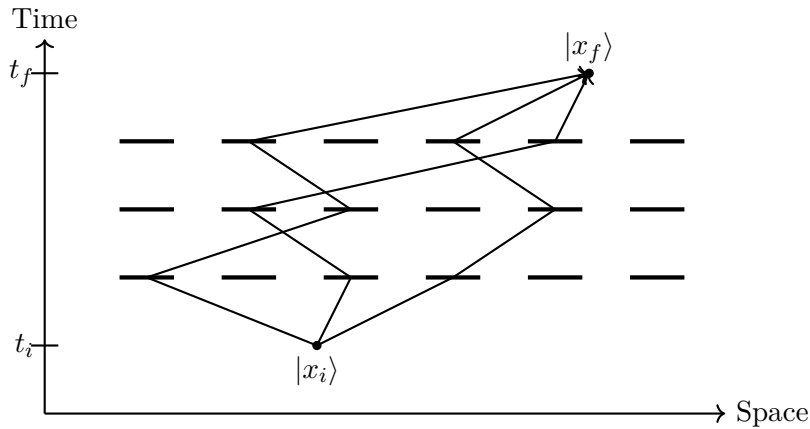
\begin{figure}[h]
		\centering
		\begin{tikzpicture}[scale=0.9]
			\draw[thick, ->] (0,0) -- (10,0) node[right] {Space};
			\draw[thick, ->] (0,0) -- (0,5.5) node[above] {Time};
			\draw[thick] (-0.2, 1) -- (0.2, 1) node[left=5pt] {$t_i$};
			\draw[thick] (-0.2, 5) -- (0.2, 5) node[left=5pt] {$t_f$};
			
			\filldraw (4, 1) circle (0.06) node[below] {$|x_i\rangle$};
			\filldraw (8, 5) circle (0.06) node[above] {$|x_f\rangle$};
			
			\foreach \y in {2, 3, 4} {
				\foreach \x in {1.5, 3, 4.5, 6, 7.5, 9} {
					\draw[ultra thick] (\x-0.4, \y) -- (\x+0.4, \y);
				}
			}
			
			\draw[->, thick] (4,1) -- (1.5,2) -- (4.5,3) -- (3,4) -- (8,5);
			\draw[->, thick] (4,1) -- (6,2) -- (7.5,3) -- (6,4) -- (8,5);
			\draw[->, thick] (4,1) -- (4.5,2) -- (3,3) -- (7.5,4) -- (8,5);
		\end{tikzpicture}
		\caption{Multiple intermediate time slices approaching the continuum limit.}
	\end{figure}
	
	The total amplitude is the integral over all these paths: 
	\begin{equation}
		\mathcal{A}_{A \rightarrow B} = \sum_{\text{paths}} e^{i\frac{S}{\hbar}} \longrightarrow \int [dx] e^{i\frac{S}{\hbar}}
	\end{equation}
	where $S$ is the classical action evaluated along a specific path, and $[dx]$ denotes the path measure. In principle one should sum over all possible paths, including those which are nowhere differentiable. A curious fact: the set of smooth paths carries measure zero because there are far many more discontinuous paths than continuous ones. 
	
	\textbf{So why do we only see one single classical path macroscopically?} 
	Every possible path has exactly the same weight. However, their phases differ. For macroscopic objects, $\hbar$ is tiny compared to the action $S$. This causes the phase $S/\hbar$ to oscillate rapidly for paths far from the classical path, leading to destructive interference. Only paths close to the stationary classical path constructively interfere. 
	That's also what happens with reflected light. 
	
	\begin{figure}[h]
		\centering
		\begin{tikzpicture}[scale=1]
			\draw[ultra thick] (0,0) -- (10,0); 
			\filldraw (1, 4) circle (0.06) node[above] {A};
			\filldraw (9, 4) circle (0.06) node[above] {B};
			
			\foreach \x/\lbl in {2/{c_1}, 3.5/{c_2}, 5/{c_0}, 6.5/{c_3}, 8/{c_4}} {
				\draw[->, thin] (1,4) -- (\x, 0);
				\draw[->, thin] (\x,0) -- (9, 4);
				\node[below] at (\x, -0.1) {$\lbl$};
			}
		\end{tikzpicture}
		\caption{Reflected light paths. Constructive interference primarily occurs along the path $c_0$ which minimizes travel time (Fermat's Principle).}
	\end{figure}
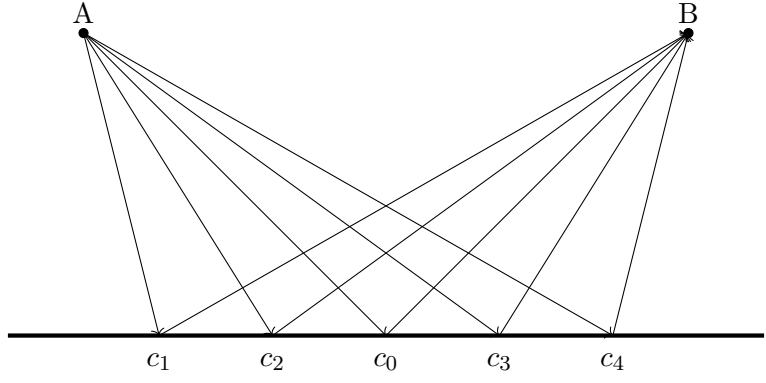
	
	\section{Deriving the Feynman Path Integral}
	We start with the standard quantum mechanical transition amplitude from an initial state $|x_a\rangle$ at time $t_a$ to a final state $|x_b\rangle$ at time $t_b$. The evolution operator uses the Hamiltonian $\hat{H}$: 
	\begin{equation}
		\mathcal{A}_{a \rightarrow b} = \langle x_b | e^{-\frac{i}{\hbar}\hat{H}(t_b - t_a)} | x_a \rangle
	\end{equation}
	where the Hamiltonian is the sum of kinetic and potential operators, satisfying $[\hat{x}, \hat{p}] = i\hbar$: 
	\begin{equation}
		\hat{H} = \frac{\hat{p}^2}{2m} + V(\hat{x})
	\end{equation}
	
	\subsection{Time Slicing and Inserting Unit Operators}
	Take $t_a = 0$ and $t_b = t$. Slice the time $[0, t]$ into $n$ pieces, $t = n\epsilon$. 
	\begin{equation}
		e^{-\frac{i}{\hbar}\hat{H}t} = e^{-\frac{i}{\hbar}\hat{H}\epsilon} e^{-\frac{i}{\hbar}\hat{H}\epsilon} \dots e^{-\frac{i}{\hbar}\hat{H}\epsilon}
	\end{equation}
	Insert the unit operator at every point: $\hat{1} = \int dx_i |x_i\rangle\langle x_i|$. 
	This yields:
	\begin{align}
		\mathcal{A}_{a \rightarrow b} &= \mathcal{A}_{a \rightarrow 1}\mathcal{A}_{1 \rightarrow 2} \dots \mathcal{A}_{i \rightarrow i+1} \dots \mathcal{A}_{n-1 \rightarrow b} \nonumber \\
		&= \int dx_1 \dots \int dx_{n-1} \langle x_b|e^{-\frac{i}{\hbar}\hat{H}\epsilon}|x_{n-1}\rangle \dots \langle x_1|e^{-\frac{i}{\hbar}\hat{H}\epsilon}|x_a\rangle
	\end{align}
	
	\subsection{The Short-Time Propagator}
	Find an expression for $\mathcal{A}_{i \rightarrow i+1} \equiv \langle x_{i+1} | e^{-\frac{i}{\hbar}\hat{H}\epsilon} | x_i \rangle$. 
	\begin{equation}
		\langle x_{i+1} | e^{-\frac{i}{\hbar}\hat{H}\epsilon} | x_i \rangle = \langle x_{i+1} | e^{-\frac{i}{\hbar}\frac{\hat{p}^2}{2m}\epsilon} e^{-\frac{i}{\hbar}V(\hat{x})\epsilon} | x_i \rangle
	\end{equation}
	Insert the momentum identity operator $\hat{1} = \int dp |p\rangle\langle p|$ here: 
	\begin{equation}
		\langle x_{i+1} | e^{-\frac{i}{\hbar}\hat{H}\epsilon} | x_i \rangle = \int_{-\infty}^{\infty} dp \, e^{-\frac{i}{\hbar}\frac{p^2}{2m}\epsilon} \langle x_{i+1}|p\rangle \langle p|x_i\rangle e^{-\frac{i}{\hbar}V(x_i)\epsilon}
	\end{equation}
	Using the explicit form of the momentum eigenstates: 
	\begin{equation}
		\langle x_{i+1}|p\rangle = \frac{1}{\sqrt{2\pi\hbar}} e^{\frac{i}{\hbar}px_{i+1}}, \quad \langle p|x_i\rangle = \frac{1}{\sqrt{2\pi\hbar}} e^{-\frac{i}{\hbar}px_i}
	\end{equation}
	We get:
	\begin{equation}
		= e^{-\frac{i}{\hbar}V(x_i)\epsilon} \int_{-\infty}^{\infty} \frac{dp}{2\pi\hbar} e^{-\frac{i}{\hbar}\frac{p^2}{2m}\epsilon} e^{\frac{i}{\hbar}p(x_{i+1} - x_i)}
	\end{equation}
	Completing the square in the exponent by setting $\dot{x} \approx \frac{x_{i+1}-x_i}{\epsilon}$: 
	\begin{equation}
		= e^{\frac{i}{2\hbar} m \left(\frac{x_{i+1}-x_i}{\epsilon}\right)^2 \epsilon} e^{-\frac{i}{\hbar}V(x_i)\epsilon} \int_{-\infty}^{\infty} \frac{dp}{2\pi\hbar} e^{-\frac{i\epsilon}{2\hbar m} \left( p - m\frac{x_{i+1}-x_i}{\epsilon} \right)^2}
	\end{equation}
	
	We substitute $s = p - m\frac{x_{i+1}-x_i}{\epsilon} = p - m\dot{x}$. The expression becomes: 
	\begin{equation}
		e^{i\frac{\epsilon}{\hbar} \left( \frac{1}{2}m\dot{x}^2 - V(x_i) \right)} \int_{-\infty}^{\infty} \frac{ds}{2\pi\hbar} e^{-\frac{i\epsilon}{2\hbar m}s^2}
	\end{equation}
	Use the standard Gaussian integral $\int_{-\infty}^{\infty} dx \, e^{-iax^2} = \sqrt{\frac{\pi}{ia}}$. 
	Applying this with $a = \frac{\epsilon}{2\hbar m}$: 
	\begin{equation}
		= \sqrt{\frac{m}{2\pi i \hbar \epsilon}} e^{\frac{i}{\hbar}\epsilon \mathcal{L}}
	\end{equation}
	where $\mathcal{L} = \frac{1}{2}m\dot{x}^2 - V(x_i)$ is the classical Lagrangian. 
	
	Substituting this back into the full amplitude and taking the limit $n \rightarrow \infty$, the discrete sum over $i$ becomes a continuous integral: 
	\begin{equation}
		\mathcal{A}_{a \rightarrow b} = \lim_{n \rightarrow \infty} \left[ \frac{m}{2\pi i \hbar \epsilon} \right]^{\frac{n}{2}} \int dx_1 \dots \int dx_{n-1} \exp\left( \frac{i}{\hbar} \sum_{j=0}^{n-1} \epsilon \mathcal{L}_j \right)
	\end{equation}
	Which formally transitions into the continuous Path Integral expression: 
	\begin{equation}
		\mathcal{A}_{a \rightarrow b} = \int \mathcal{D}x(t) \, e^{\frac{i}{\hbar} S[x(t)]}
	\end{equation}
	
	\section{Computation in Simple Cases}
	The path integral framework derived above allows for exact analytical evaluations for physical systems with actions at most quadratic in the fields. 
	
	\subsection{The Free Particle}
	For a free particle of mass $m$, the potential $V(x) = 0$. The Lagrangian is simply the kinetic energy: 
	\begin{equation}
		\mathcal{L} = \frac{1}{2}m\dot{x}^2
	\end{equation}
	The classical path $x_{cl}(t)$ connecting $x(t_a) = x_a$ to $x(t_b) = x_b$ is a straight line, representing constant velocity $v = \frac{x_b - x_a}{T}$, where $T = t_b - t_a$. 
	The classical action evaluated along this path is: 
	\begin{equation}
		S_{cl} = \int_{t_a}^{t_b} \frac{1}{2}m \left( \frac{x_b - x_a}{T} \right)^2 dt = \frac{m(x_b - x_a)^2}{2T}
	\end{equation}
	Since the action is purely quadratic, quantum fluctuations around the classical path can be isolated. 
	By writing any path as $x(t) = x_{cl}(t) + y(t)$, where $y(t_a) = y(t_b) = 0$, the path integral factors into: 
	\begin{equation}
		\mathcal{A}_{a \rightarrow b} = e^{\frac{i}{\hbar}S_{cl}} \int \mathcal{D}y(t) \, e^{\frac{i}{\hbar} \int_{t_a}^{t_b} \frac{1}{2}m\dot{y}^2 dt} = A(T) e^{\frac{i}{\hbar}S_{cl}}
	\end{equation}
	Evaluating the fluctuation determinant $A(T)$ gives the final known result for the free particle propagator: 
	\begin{equation}
		K_{\text{free}}(x_b, t_b; x_a, t_a) = \sqrt{\frac{m}{2\pi i \hbar T}} \exp\left( \frac{im(x_b - x_a)^2}{2\hbar T} \right)
	\end{equation}
	
	\subsection{The Quantum Harmonic Oscillator}
	For a particle subject to a harmonic potential, the Lagrangian is $\mathcal{L} = \frac{1}{2}m\dot{x}^2 - \frac{1}{2}m\omega^2 x^2$. 
	The classical path $x_{cl}(t)$ obeys the equation of motion $\ddot{x} + \omega^2 x = 0$. 
	Solving this yields the classical action: 
	\begin{equation}
		S_{cl} = \frac{m\omega}{2\sin(\omega T)} \left[ (x_a^2 + x_b^2)\cos(\omega T) - 2x_a x_b \right]
	\end{equation}
	Similar to the free particle, the path integral separates into a classical exponential and a fluctuation factor $F(T)$: 
	\begin{equation}
		\mathcal{A}_{a \rightarrow b} = F(T) e^{\frac{i}{\hbar} S_{cl}}
	\end{equation}
	The fluctuation integral is independent of $x_a$ and $x_b$, evaluating only paths $y(t)$ with zero endpoints: 
	\begin{equation}
		F(T) = \int \mathcal{D}y(t) \, e^{\frac{i}{\hbar} \int_0^T \left( \frac{1}{2}m\dot{y}^2 - \frac{1}{2}m\omega^2 y^2 \right) dt} = \sqrt{\frac{m\omega}{2\pi i \hbar \sin(\omega T)}}
	\end{equation}
	Thus, the complete transition amplitude for the harmonic oscillator is: 
	\begin{equation}
		G_{\text{osc}}(x_b, t_b; x_a, t_a) = \sqrt{\frac{m\omega}{2\pi i \hbar \sin(\omega T)}} \exp\left\{ \frac{im\omega}{2\hbar \sin(\omega T)} \left[ (x_a^2 + x_b^2)\cos(\omega T) - 2x_a x_b \right] \right\}.
	\end{equation}
	
	\section{Review of Functional Analysis}
	Before delving further into path integrals, we review the fundamental concepts of functional analysis. 
	While a standard function $f(x)$ maps a variable to a value, a functional $F[g]$ maps an entire function $g(x)$ to a numerical value. 
	\begin{equation}
		F[g] = \int_a^b dx \, L(x, g(x), g'(x), g''(x), \dots)
	\end{equation}
	The standard derivative for a function is $\frac{df(x)}{dx} = \lim_{\epsilon \rightarrow 0} \frac{f(x+\epsilon) - f(x)}{\epsilon}$. 
	Similarly, the functional derivative with respect to a function $g(x)$ is evaluated using the Dirac delta function $\delta(x-y)$: 
	\begin{equation}
		\frac{\delta F[g(y)]}{\delta g(x)} = \lim_{\epsilon \rightarrow 0} \frac{F[g(y) + \epsilon \delta(x-y)] - F[g(y)]}{\epsilon}
	\end{equation}
	For example, just as $f(x) = x^2 \rightarrow f' = 2x$, the corresponding functional derivative for $F[g(x)] = g(x)^2$ yields $F' = 2\delta(x-y)$. 
	
	\subsection{Functional Taylor Expansion}
	The usual Taylor expansion for a function is $f(x + \delta x) = f(x) + \delta x f'(x) + \frac{1}{2} \delta x^2 f''(x) + \dots$. 
	Similarly, there is a Taylor expansion for functionals: 
	\begin{equation}
		F[g(x) + \delta g(x)] = F[g(x)] + \int_a^b dx' \frac{\delta F[g(x)]}{\delta g(x')} \delta g(x') + \frac{1}{2} \int_a^b dx' \int_a^b dx'' \frac{\delta^2 F[g(x)]}{\delta g(x')\delta g(x'')} \delta g(x')\delta g(x'') + \dots
	\end{equation}
	
	\section{Real Time and Imaginary Time Path Integrals}
	The transition amplitude in real time is given by the sum over all paths: 
	\begin{equation}
		\mathcal{A}_{a \rightarrow b} = \sum_{\text{paths}} e^{i\frac{S}{\hbar}} \longrightarrow \int [dx] e^{i\frac{S[x]}{\hbar}}
	\end{equation}
	where the measure $[dx] = \mathcal{N} \lim_{n \rightarrow \infty} \prod_{i=1}^{n-1} dx_i$ and the action is $S[x] = \int_0^t dt \left(\frac{1}{2}m\dot{x}^2 - V(x)\right)$. 
	
	\begin{figure}[h]
		\centering
		\includegraphics[width=0.6\textwidth]{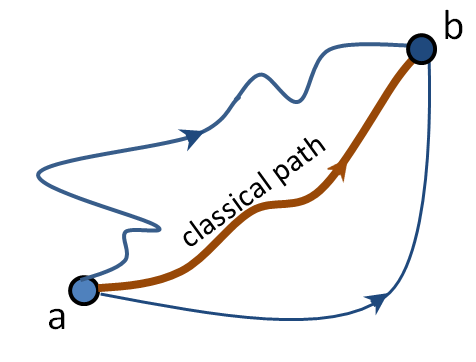}
		\caption{In the real-time path integral, the transition amplitude from point $a$ to point $b$ is dominated by the classical path due to constructive interference, while wildly oscillating non-classical paths destructively interfere.}
		\label{fig:real_paths}
	\end{figure}
	
	\subsection{Imaginary Time Path Integral}
	The problem: For $\hbar \rightarrow 0$, the phase factor $e^{iS/\hbar}$ oscillates wildly, making exact analytical and numerical evaluation incredibly difficult. 
	To cure this oscillatory behavior, we perform a Wick rotation to imaginary time. We put $t = i\tau$. 
	
	Under this transformation, the kinetic term becomes negative: 
	\begin{equation}
		\left(\frac{dx}{dt}\right)^2 = -\left(\frac{dx}{d\tau}\right)^2 \equiv -\dot{x}^2
	\end{equation}
	(where the dot now denotes differentiation with respect to $\tau$). The time integration measure transforms as $\frac{i}{\hbar} \int dt = \frac{1}{\hbar} \int d\tau$. 
	
	Substituting this into the action gives the \textbf{Euclidean action}: 
	\begin{equation}
		S[x] = \int_{\tau_a}^{\tau_b} d\tau \left( \frac{1}{2}m\dot{x}^2 + V(x) \right)
	\end{equation}
	Notice that the potential term $V(x)$ has effectively flipped its sign. 
	The transition amplitude now takes the form of a decaying exponential rather than an oscillatory phase: 
	\begin{equation}
		\mathcal{A}_{A \rightarrow B} = \int [dx] \exp\left(-\frac{S[x]}{\hbar}\right)
	\end{equation}
	
	\subsection{Some Low Hanging Fruit}
	Let us evaluate the derivative of the ``energy'' $E$ along the classical path $x_c(\tau)$: 
	\begin{equation}
		\frac{dE}{d\tau} \equiv \frac{d}{d\tau}\left( -\frac{1}{2}m\dot{x}_c^2 + V(x_c) \right) = -m\dot{x}_c\ddot{x}_c + V'(x_c)\dot{x}_c = (-m\ddot{x}_c + V'(x_c))\dot{x}_c = 0
	\end{equation}
	Hence, the ``energy'' $E$ is conserved on $x_c(\tau)$ (but not away from it!). 
	
	\newpage
	\part{Fluctuation Determinants and the Gel'fand-Yaglom Theorem}
	
	\section{Separating the Action into Classical and Fluctuation Parts}
	We have factored the transition amplitude into a classical part and a fluctuation part: 
	\begin{equation}		
		\mathcal{A}_{a \rightarrow b} = \langle x_b | e^{-\frac{i}{\hbar}\hat{H}(t_b - t_a)} | x_a \rangle=\int [dx] e^{-\frac{1}{\hbar}S[x]} =  \mathcal{A}_{\text{classical}}^{(0)} \times \mathcal{A}_{\text{fluctuation}}^{(2)},
	\end{equation}
	where,
	\begin{equation}
		\mathcal{A}_{\text{classical}}^{(0)}= e^{-\frac{1}{\hbar}S^{(0)}[x_c]},\quad \mathcal{A}_{\text{fluctuation}}^{(2)}= \int [d\xi] e^{-\frac{1}{\hbar}S^{(2)}[\xi]}.
	\end{equation}
	A general arbitrary path can be written as $x(\tau) = x_c(\tau) + \xi(\tau)$ with boundary conditions $\xi(\tau_a) = \xi(\tau_b) = 0$, so we can set $[dx] = [d\xi]$. 
	The above followed because we could separate the actions: $S_{\text{total}} = S_{\text{classical}}^{(0)} + S_{\text{fluctuation}}^{(2)}$. 
	
	\section{The Fluctuation Determinant}
	How do we compute the fluctuation action $S^{(2)}$? 
	\begin{equation}
		S^{(2)} = \frac{1}{2} \int_{\tau_a}^{\tau_b} d\tau \, \xi(\tau) \left(-m\frac{d^2}{d\tau^2} + V''(x_c)\right)\xi(\tau)
	\end{equation}
	Let's expand the fluctuation $\xi(\tau)$ in the complete orthonormal basis $\{\phi_n(\tau)\}$, defined by the eigenvalue equation: 
	\begin{equation}
		\left(-m\frac{d^2}{d\tau^2} + V''(x_c)\right)\phi_n(\tau) = \lambda_n \phi_n(\tau)
	\end{equation}
	with boundary conditions $\phi_n(\tau_a) = \phi_n(\tau_b) = 0$. 
	The path can be expressed as $x(\tau) = x_c(\tau) + \xi(\tau) = x_c(\tau) + \sum_{n=0}^{\infty} c_n \phi_n(\tau)$. 
	Substituting this expansion into $S^{(2)}$ orthogonalizes the integral: 
	\begin{equation}
		S^{(2)} = \frac{1}{2} \sum_{n=0}^{\infty} \lambda_n c_n^2
	\end{equation}
	We evaluate the path integral over the fluctuations: $\int [d\xi] e^{-\frac{1}{\hbar}S^{(2)}(\xi)}$ 
	where the measure $[dx] = [d\xi]$ is written in terms of the expansion coefficients $c_n$: 
	\begin{equation}
		[dx] = \mathcal{N} \prod_{n=0}^{\infty} \frac{dc_n}{\sqrt{2\pi\hbar}}
	\end{equation}
	Here $\mathcal{N}$ is the normalization constant which we will eventually need to figure out. 
	Because integrating over paths is equivalent to integrating over the coefficients $c_n$, the integral becomes a product of infinite Gaussian integrals: 
	\begin{equation}
		\int [dx] e^{-\frac{1}{\hbar}S^{(2)}} = \mathcal{N} \prod_{n=0}^{\infty} \int_{-\infty}^{\infty}\frac{dc_n}{\sqrt{2\pi\hbar}} e^{-\frac{1}{2\hbar}\lambda_n c_n^2} = \mathcal{N} \prod_{n=0}^{\infty} \frac{1}{\sqrt{\lambda_n}}
	\end{equation}
	Recognizing the product of eigenvalues $\prod \lambda_n$ as the determinant of the operator $S'' \equiv -m\frac{d^2}{d\tau^2} + V''(x_c)$, we arrive at the final beautifully simple form for the transition amplitude: 
	\begin{equation}
		\mathcal{A}_{a \rightarrow b} = e^{-\frac{1}{\hbar}S^{(0)}[x_c]} \frac{\mathcal{N}}{\sqrt{\det S''}}
	\end{equation}
	
	Notice that the normalization constant $\mathcal{N}$ involves an infinite product that fundamentally diverges. 
	How do we determine it? 
	In practice, we rarely compute $\mathcal{N}$ directly. 
	Instead, we normalize our transition amplitude against a known, exactly solvable reference system---typically the free particle amplitude $\mathcal{A}_{\text{free}}$ or the simple harmonic oscillator $\mathcal{A}_{\text{sho}}$. 
	By taking a ratio the divergent constant $\mathcal{N}$ neatly cancels out, leaving only the ratio of the functional determinants: 
	\begin{equation}	
		\frac{\mathcal{A}}{\mathcal{A}_{\text{free}}} = \frac{ e^{-\frac{1}{\hbar}S^{(0)}}}{e^{-\frac{1}{\hbar}S^{(0)}_{\text{free}}}} \left[ \frac{\det\left(-m\frac{d^2}{d\tau^2} + V''(x_c)\right)}{\det\left(-m\frac{d^2}{d\tau^2}\right)} \right]^{-1/2}
		\label{ratio}
	\end{equation}	
	
	\subsection{Exercise: Comparing a SHO and Free Particle}
	\textbf{Problem:} Consider a particle moving in Euclidean time from $(x_a, -T/2)$ to $(x_b, T/2)$. For a simple harmonic oscillator (SHO) and a free particle, first evaluate the classical actions and specify the required fluctuation operators in the above equation. 
	
	\noindent\textbf{Solution:}
	The Euclidean action is $S_E = \int_{-T/2}^{T/2} d\tau \left( \frac{1}{2}m\dot{x}^2 + V(x) \right)$, evaluated along the classical path $x_c(\tau)$ that satisfies the given boundary conditions. 
	
	\noindent\textbf{1. The Free Particle:}
	\begin{itemize}
		\item The potential is $V(x) = 0$, hence the fluctuation operator is simply $-m \frac{d^2}{d\tau^2}$. 
		\item The classical EOM is $\ddot{x}_c = 0$, yielding the linear path $x_c(\tau) = x_a + \frac{x_b - x_a}{T}(\tau +T/2)$. 
		\item Integrating the kinetic energy along this path gives $S^{(0)}_{\text{free}} = \frac{m(x_b - x_a)^2}{2T}$. 
	\end{itemize}
	
	\noindent\textbf{2. The Simple Harmonic Oscillator:}
	\begin{itemize}
		\item For $V(x) = \frac{1}{2}m\omega^2 x^2, \; V''(x_c) = m\omega^2$, and the fluctuation operator is $-m \frac{d^2}{d\tau^2} + m\omega^2$. 
		\item The EOM is $\ddot{x}_c - \omega^2 x_c = 0$ gives:
		$x_c(\tau) = \frac{x_b + x_a}{2 \cosh\left(\frac{\omega T}{2}\right)} \cosh(\omega \tau) + \frac{x_b - x_a}{2 \sinh\left(\frac{\omega T}{2}\right)} \sinh(\omega \tau).$ 
		\item The action along this trajectory is:
		$S^{(0)}_{\text{sho}} = \frac{m\omega}{2\sinh(\omega T)} \left[ (x_a^2 + x_b^2)\cosh(\omega T) - 2x_a x_b \right].$ 
	\end{itemize}
	
	With the classical actions having now been evaluated, the ratio in Eq.\ref{ratio} becomes: 
	\begin{equation}
		\frac{ e^{-\frac{1}{\hbar}S^{(0)}_{\text{sho}}}}{e^{-\frac{1}{\hbar}S^{(0)}_{\text{free}}}} \times \left[ \frac{\det \left( - \frac{d^2}{d\tau^2} + \omega^2 \right)}{\det \left( - \frac{d^2}{d\tau^2} \right)} \right]^{-1/2}
	\end{equation}
	We will soon see how to evaluate the above ratio of determinants. 
	Certainly, we do know the full eigenvalue spectrum of these two elementary differential operators. 
	But, in the case of a general 1-D potential, this could be an impossibly difficult task. 
	Fortunately, from functional analysis we have at our disposal the powerful Gel'fand-Yaglom theorem, which reduces the entire determinant ratio to the solution of a single initial-value ordinary differential equation. 
	We now turn to this important theorem. 
	
	\section{Determinant Ratios via the Gel'fand-Yaglom Theorem}
	For simplicity put $m=1$ and suppose $W(\tau)$ is an arbitrary confining potential. 
	The eigenvalue problem with Dirichlet boundary conditions is: 
	\begin{equation}
		\left( -\frac{d^2}{d\tau^2} + W(\tau) \right) \phi_n = \lambda_n \phi_n \quad \text{with} \quad \phi_n(-T/2) = \phi_n(T/2) = 0
	\end{equation}
	
	\begin{figure}[h]
		\centering
		\includegraphics[width=0.5\textwidth]{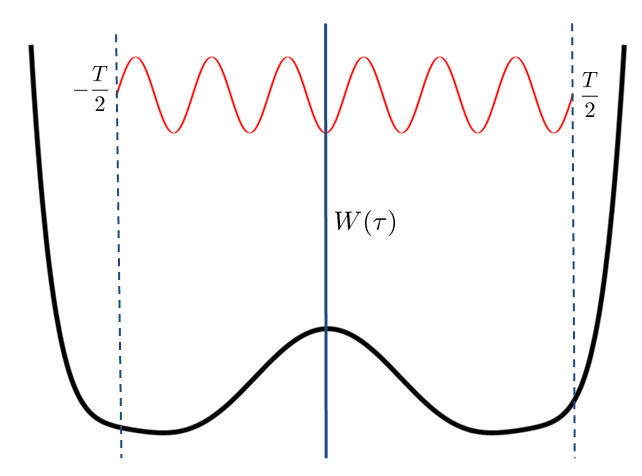}
		\caption{As the eigenvalue $\lambda$ (represented by the high-frequency wave) becomes very large, the underlying potential $W(\tau)$ becomes less dynamically relevant to the solutions.}
		\label{fig:high_energy}
	\end{figure}
	
	As the eigenvalue $\lambda$ becomes very large, the potential $W$ becomes dynamically less relevant. 
	This means that for two different potentials $W_1$ and $W_2$, their respective eigenvalues will asymptotically approach each other: $\lim_{n \to \infty} (\lambda_n^1 - \lambda_n^2) = 0$. 
	Consequently, we can define a function $F(\lambda)$ as an infinite product: 
	\begin{equation}
		F(\lambda) = \prod_{n=1}^\infty \frac{\lambda_n^1 - \lambda}{\lambda_n^2 - \lambda}
	\end{equation}
	Because of the asymptotic behavior of the eigenvalues, this infinite product is possibly convergent and well behaved. 
	Notice its limits: 
	\begin{equation}
		\lim_{\lambda \to 0} F(\lambda) = \prod_{n=1}^\infty \frac{\lambda_n^1}{\lambda_n^2} = \frac{\det_1}{\det_2} \quad \text{and} \quad \lim_{\lambda \to \infty} F(\lambda) = 1
	\end{equation}
	
	Now, given any $W(\tau)$, the Cauchy (initial value) problem for any $\lambda$ is always solvable. 
	Let us define a function $\phi_\lambda(\tau)$ such that: 
	\begin{equation}
		\left( -\frac{d^2}{d\tau^2} + W(\tau) \right) \phi_\lambda = \lambda \phi_\lambda
	\end{equation}
	with initial position $\phi_\lambda(-T/2) = 0$ and initial speed $\dot{\phi}_\lambda(-T/2) = 1$. 
	
	In general, for an arbitrary $\lambda$, the ``ball'' will overshoot or undershoot the final target boundary condition at $T/2$. 
	Only when $\lambda$ equals one of the exact discrete eigenvalues $\lambda_n$ will the trajectory reach the other hilltop exactly, in which case $\phi_\lambda(T/2) = 0$. 
	
	Furthermore, every potential will give the same value for $\phi_\lambda(T/2)$ for large enough $\lambda$. 
	Therefore, we might as well choose $W_1$ to be a simple constant, such as $W_1 = \omega^2$. 
	
	Now, define a second function $G(\lambda)$ as the ratio of these endpoint values: 
	\begin{equation}
		G(\lambda) = \frac{\phi_\lambda^1(T/2)}{\phi_\lambda^2(T/2)}
	\end{equation}
	Both $F(\lambda)$ and $G(\lambda)$ have poles and zeros at exactly the same discrete values of $\lambda$. 
	They also agree in the asymptotic limit for large $|\lambda|$. 
	Because $F(\lambda)$ and $G(\lambda)$ are holomorphic functions, Liouville's theorem strictly applies, and hence they must agree everywhere! 
	
	Evaluating them at $\lambda = 0$ gives the celebrated Gel'fand-Yaglom result: 
	\begin{equation}
		\frac{\det_1}{\det_2} = \frac{\phi_{1,0}(T/2)}{\phi_{2,0}(T/2)}
	\end{equation}
	This remarkable theorem converts the intractable product of infinite eigenvalues into a straightforward initial value differential equation problem. 
	
	\subsection{Exercise: The Harmonic Oscillator Fluctuation Factor}
	\textbf{Problem:} Use the Gel'fand-Yaglom theorem to evaluate the functional determinant factor associated with the simple harmonic oscillator, given by: 
	\begin{equation*}
		\left[ \frac{\det \left( -\frac{d^2}{d\tau^2} + \omega^2 \right)}{\det \left( -\frac{d^2}{d\tau^2} \right)} \right]^{-1/2}
	\end{equation*}
	Assume the time interval is symmetric: $\tau \in [-T/2, T/2]$. 
	
	\noindent\textbf{Solution:}
	Let us identify the two operators appearing in the numerator and denominator. 
	For the simple harmonic oscillator, the operator is $O_1 = -\frac{d^2}{d\tau^2} + W_1(\tau)$, where the effective potential is a constant, $W_1(\tau) = \omega^2$. 
	For the free particle reference system, the operator is $O_2 = -\frac{d^2}{d\tau^2} + W_2(\tau)$, where $W_2(\tau) = 0$. 
	
	According to the Gel'fand-Yaglom theorem, the ratio of their determinants is equal to the ratio of the solutions of their respective homogeneous initial value problems, evaluated at the final time $+T/2$. 
	We must solve $O_i \phi_i(\tau) = 0$ subject to the Dirichlet initial conditions at the left boundary $\tau_a = -T/2$: 
	\begin{equation*}
		\phi_i(-T/2) = 0 \quad \text{and} \quad \dot{\phi}_i(-T/2) = 1
	\end{equation*}
	
	\textbf{Step 1: Evaluate the numerator function $\phi_1(\tau)$} \\
	The differential equation for the harmonic oscillator is: $\ddot{\phi}_1(\tau) - \omega^2 \phi_1(\tau) = 0$. 
	The solution that satisfies $\phi_1(-T/2)=0,\; \dot{\phi}_1(-T/2) = 1$ is, $\phi_1(\tau) = \frac{1}{\omega} \sinh\left[ \omega \left(\tau + \frac{T}{2} \right) \right]$. 
	Evaluating this function at the final time boundary $\tau = T/2$ provides the numerator for our GY ratio: 
	\begin{equation}
		\phi_1(T/2) = \frac{\sinh(\omega T)}{\omega}
	\end{equation}
	
	\textbf{Step 2: Evaluate the denominator function $\phi_2(\tau)$} \\
	The differential equation for the free particle is simply: $\ddot{\phi}_2(\tau) = 0$. 
	The solution is a straight line: $\phi_2(\tau) = \tau + \frac{T}{2}$. 
	Evaluating this function at the final time boundary $\tau = T/2$ provides the denominator for our GY ratio: 
	\begin{equation}
		\phi_2(T/2) = T
	\end{equation}
	
	\textbf{Step 3: Calculate the final fluctuation factor} \\
	By the Gel'fand-Yaglom theorem, the ratio of the functional determinants is exactly the ratio of these final endpoint values: 
	\begin{equation*}
		\frac{\det \left( -\frac{d^2}{d\tau^2} + \omega^2 \right)}{\det \left( -\frac{d^2}{d\tau^2} \right)} = \frac{\phi_1(T/2)}{\phi_2(T/2)} = \frac{\sinh(\omega T)}{\omega T}.
	\end{equation*}
	
	\newpage
	\part{Zero Modes and the Dilute Instanton Gas}
	
	\section{The Symmetric Double Well Potential}
	We now apply our imaginary time formalism to the symmetric double-well potential, $V(x) = \frac{\lambda}{4}(x^2 - a^2)^2$. 
	This potential has two degenerate minima at $x = \pm a$. 
	
	\begin{figure}[h]
		\centering
		\includegraphics[width=0.8\textwidth]{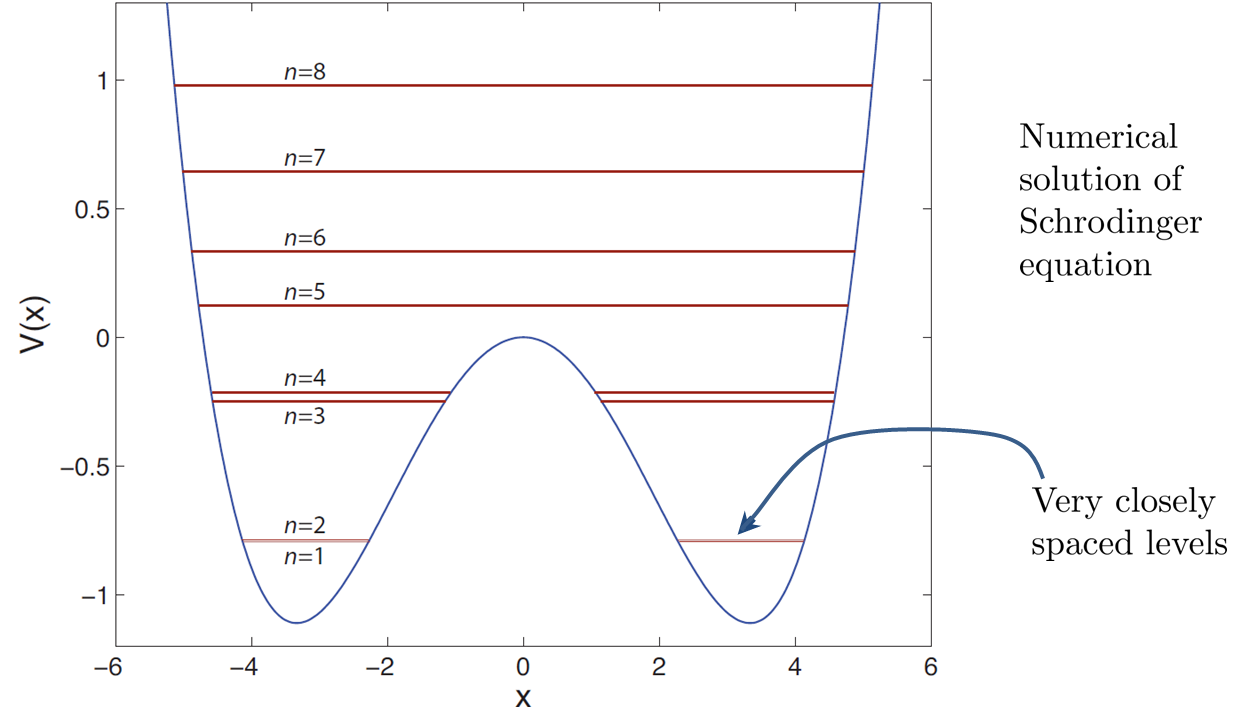}
		\caption{Numerical solution of the Schrödinger equation for the symmetric double-well potential reveals pairs of extremely closely spaced energy levels. This fine structure cannot be explained by classical mechanics alone.}
		\label{fig:levels}
	\end{figure}
	
	In real time, a particle with energy $E < V(0)$ localized at $-a$ cannot classically reach $+a$. 
	Classically, a particle would stay in the left or right minimum forever. 
	However, quantum mechanically, a state perfectly localized in one well is not a true eigenstate of the full Hamiltonian due to tunneling effects. 
	
	\begin{figure}[h]
		\centering
		\begin{minipage}{0.48\textwidth}
			\centering \includegraphics[width=\linewidth]{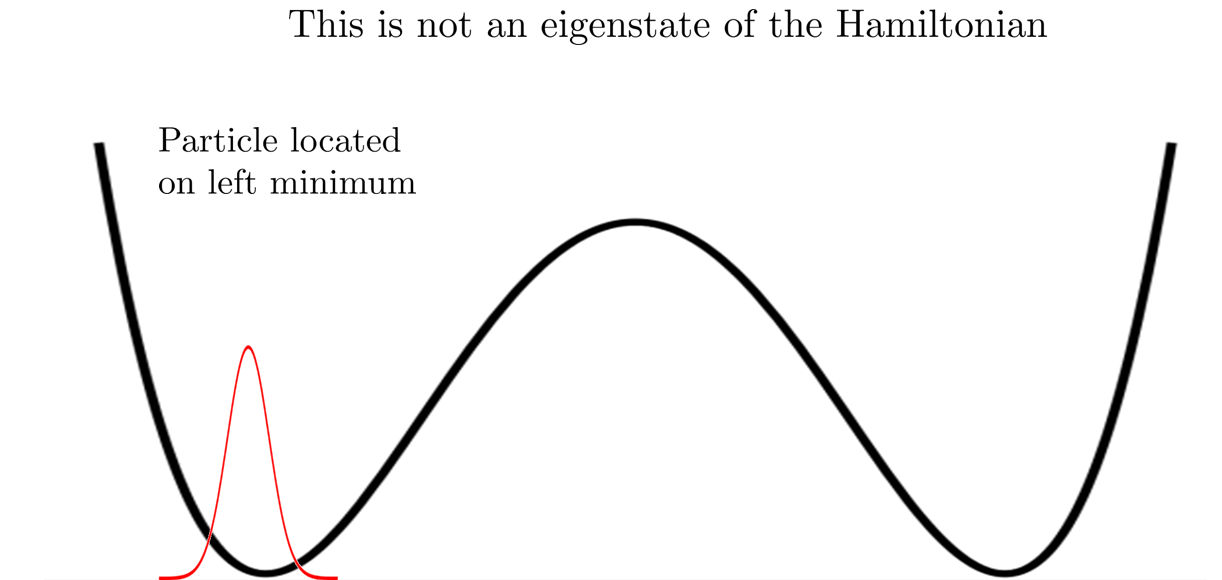}
			\caption{Particle localized in the left minimum. This is not an eigenstate of the Hamiltonian.}
			\label{fig:left_not_eigen}
		\end{minipage}\hfill
		\begin{minipage}{0.48\textwidth}
			\centering \includegraphics[width=\linewidth]{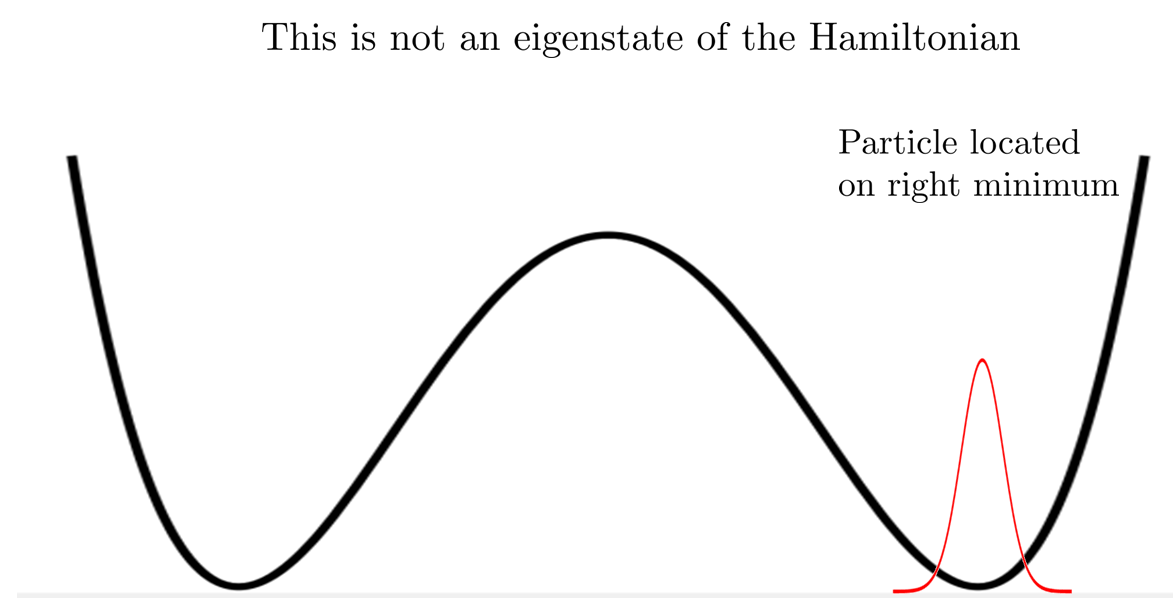}
			\caption{Particle localized in the right minimum. This is also not a true eigenstate.}
			\label{fig:right_not_eigen}
		\end{minipage}
	\end{figure}
	
	Recall that the Wick rotation to imaginary time ($t = i\tau$) effectively inverts the potential: $V(x) \rightarrow -V(x)$. 
	In the inverted potential, the former minima at $\pm a$ become two hilltops, and the central barrier at $x=0$ becomes a deep valley. 
	
	\begin{figure}[h]
		\centering
		\includegraphics[width=0.8\textwidth]{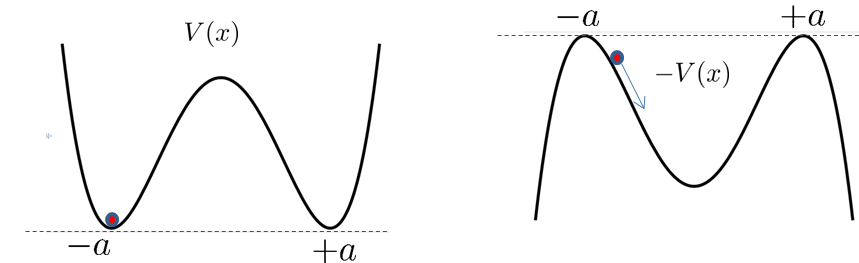}
		\caption{The original symmetric double well potential $V(x)$ (left) and the inverted potential $-V(x)$ (right) obtained via Wick rotation. The particle now classically rolls down from one hilltop and glides to the opposite hilltop.}
		\label{fig:inverted_V}
	\end{figure}
	
	An instanton is simply the classical trajectory $x_c(\tau)$ of a particle rolling from rest at the top of the left hill ($x = -a$ at $\tau = -\infty$), crossing the valley, and coming to rest exactly at the top of the right hill ($x = a$ at $\tau = +\infty$). 
	
	Let us find the exact analytical form of this trajectory. 
	Since the particle starts at rest at the hilltop where the inverted potential is zero, the conserved ``energy'' in imaginary time is zero: 
	\begin{equation}
		E = \frac{1}{2}m\dot{x}_c^2 - V(x_c) = 0 \implies \dot{x}_c = \sqrt{\frac{2}{m}V(x_c)}
	\end{equation}
	Substituting the double-well potential $V(x) = \frac{m\omega^2}{8a^2}(x^2-a^2)^2$ (where $\omega$ is the oscillator frequency at the potential minima), we obtain a separable first-order differential equation. 
	Integrating this yields the classic ``kink'' solution: 
	\begin{equation}
		x_c(\tau) = a \tanh\left( \frac{\omega}{2} (\tau - \tau_0) \right)
	\end{equation}
	
	\begin{figure}[h]
		\centering
		\includegraphics[width=0.6\textwidth]{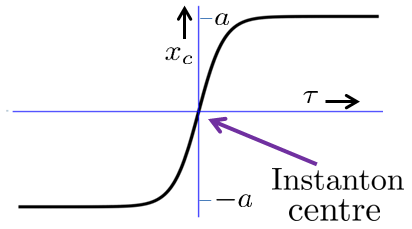}
		\caption{The classical instanton profile $x_c(\tau)$ represents a localized kink connecting $-a$ to $+a$, centered at $\tau_0$.}
		\label{fig:instanton_kink}
	\end{figure}
	
	The parameter $\tau_0$ marks the center of the instanton (the moment the particle crosses $x=0$). 
	This translation-invariant parameter $\tau_0$ represents a zero-mode in the fluctuation spectrum. 
	
	When we compute the fluctuation determinant $\det S''$ around this kink, the second derivative of the potential, $V''(x_c)$, naturally produces a Pöschl-Teller potential shape. 
	This remarkable feature ensures that the spectrum of quantum fluctuations around the instanton is exactly solvable. 
	
	Because the imaginary time equation of motion $m\ddot{x}_c = V'(x_c)$ is invariant under time reversal $\tau \rightarrow -\tau$, there exists a second, equally valid classical trajectory. 
	This is the anti-instanton, representing the tunneling event from $+a$ back to $-a$: 
	\begin{equation}
		\bar{x}_c(\tau) = -a \tanh\left( \frac{\omega}{2} (\tau - \tau_0) \right)
	\end{equation}
	
	\subsection{Action for One Instanton}
	The classical Euclidean action $S_0$ for a single instanton (or anti-instanton) is finite. 
	Using the zero-energy condition $\frac{1}{2}m\dot{x}_c^2 = V(x_c)$, the action simplifies to: 
	\begin{equation}
		S_0 = \int_{-\infty}^{\infty} d\tau \left( \frac{1}{2}m\dot{x}_c^2 + V(x_c) \right) = \int_{-\infty}^{\infty} d\tau \, m\dot{x}_c^2 = \int_{-a}^{a} dx \sqrt{2m V(x)}
	\end{equation}
	Evaluating this spatial integral for the specific shape of the double well yields exactly: $S_0 = \frac{2}{3}m\omega a^2$. 
	For simplicity going forward, we will label this single instanton action $S_1$. 
	
	\section{The Zero Mode Problem}
	We found that the path integral over the fluctuations evaluates to an infinite product of Gaussian integrals: 
	\begin{equation}
		\int [d\xi] e^{-\frac{1}{\hbar}S^{(2)}} = \mathcal{N} \prod_{n=0}^{\infty} \frac{1}{\sqrt{\lambda_n}} = \frac{\mathcal{N}}{\sqrt{\det S''}}
	\end{equation}
	The straightforward determinant calculation assumes all eigenvalues are strictly positive. 
	But what if some eigenvalue $\lambda_n$ vanishes? 
	There is no problem unless the lowest eigenvalue $\lambda_0 = 0$. 
	If $\lambda_0 = 0$, the corresponding Gaussian integral diverges: 
	\begin{equation}
		\int_{-\infty}^{\infty} \frac{dc_0}{\sqrt{2\pi\hbar}} e^{-0} = \infty
	\end{equation}
	This infinity renders the naïve determinant formally infinite, collapsing our transition amplitude. 
	It is not a mathematical artifact; instead it owes its existence to a profound physical symmetry! 
	
	\textbf{Step 1:} Suppose the instanton profile $x_c(\tau)$ is a valid solution to the imaginary-time equation of motion: $\frac{d^2}{d\tau^2} x_c(\tau) = V'(x_c(\tau))$. 
	Because the Lagrangian has no explicit time dependence, any time-shifted trajectory $x_c(\tau - \tau')$ is an equally good solution (time translational invariance). 
	
	\textbf{Step 2:} Let us differentiate the equation of motion with respect to $\tau$: 
	\begin{equation}
		\frac{d}{d\tau} \left( -\frac{d^2 x_c}{d\tau^2} + V'(x_c) \right) = \left( -\frac{d^2}{d\tau^2} + V''(x_c) \right) \frac{dx_c}{d\tau} = 0
	\end{equation}
	This directly proves that the derivative of the classical path, $\dot{x}_c(\tau)$, is an exact eigenfunction of our fluctuation operator $S'' \equiv -\frac{d^2}{d\tau^2} + V''(x_c)$ with a zero eigenvalue ($\lambda_0 = 0$). 
	
	Note that $\dot{x}_c(\tau)$ strictly obeys the correct boundary conditions, i.e., $\dot{x}_c(-T/2) = \dot{x}_c(+T/2) = 0$, because the instanton comes to a complete rest in the vacua at the integration boundaries. 
	
	\textbf{Step 3:} To use this zero mode in our basis expansion, the zero mode wavefunction must be properly normalized. We define: 
	\begin{equation}
		\phi_0(\tau) \equiv C \frac{dx_c}{d\tau}
	\end{equation}
	where the normalization constant $C$ ensures $\int \vert{}\phi_0\vert{}^2 d\tau = 1$. 
	Evaluating this integral gives: 
	\begin{equation}
		C = \left[ \int_{-T/2}^{T/2} d\tau \left( \frac{dx_c}{d\tau} \right)^2 \right]^{-1/2} = \left[ \int_{-a}^{+a} dx \sqrt{2V(x_c)} \right]^{-1/2} = S_1^{-1/2}
	\end{equation}
	Here, $S_1$ uniquely identifies the single instanton action. 
	
	\section{Assigning the Proper Integration Measure}
	To bypass the infinity associated with integrating over $dc_0$, we can choose our integration variables differently. 
	Recall that we had earlier expanded around the classical path as $x(\tau) = x_c(\tau) + \sum_{n=0}^{\infty} c_n \phi_n(\tau)$. 
	
	Instead of expanding around a fixed instanton and accumulating an infinite fluctuation $c_0$, we parameterize the path using the instanton's central position $\tau'$: 
	\begin{equation}
		x(\tau - \tau') = x_c(\tau) + \xi(\tau) = x_c(\tau) + \sum_{n=0}^{\infty} c_n \phi_n(\tau)
	\end{equation}
	Sliding the instanton along this path (changing $\tau'$) requires absolutely no effort (no change in action), which is precisely why the zero-mode eigenvalue $\lambda_0$ is zero. 
	
	To find the transformation between $c_0$ and $\tau'$, we operate on the expansion with $\int_{-T/2}^{T/2} d\tau \, \phi_0(\tau)$: 
	\begin{equation}
		\int_{-T/2}^{T/2} d\tau \, \phi_0(\tau) x(\tau - \tau') = \int_{-T/2}^{T/2} d\tau \, \phi_0(\tau) x_c(\tau) + c_0
	\end{equation}
	
	Note that by substituting $\phi_0(\tau) = C \dot{x}_c(\tau)$, the first term on the right hand side vanishes because it is a total derivative of a bounded function evaluated at the extrema: 
	\begin{equation}
		C \int_{-T/2}^{T/2} d\tau \, \dot{x}_c(\tau) x_c(\tau) = C \int_{-T/2}^{T/2} d\tau \frac{1}{2} \frac{d}{d\tau} x_c^2 = 0
	\end{equation}
	
	This isolation gives us $c_0$ as a function of the instanton position $\tau'$ for different choices of paths: 
	\begin{equation}
		c_0(\tau') = C \int_{-T/2}^{T/2} d\tau \, \dot{x}_c(\tau) x(\tau - \tau')
	\end{equation}
	
	We now differentiate $c_0$ with respect to $\tau'$ to find the Jacobian measure mapping: 
	\begin{align}
		\frac{dc_0(\tau')}{d\tau'} &= C \int_{-T/2}^{T/2} d\tau \, \dot{x}_c(\tau) \frac{d}{d\tau'} x(\tau - \tau') \nonumber \\
		&= -C \int_{-T/2}^{T/2} d\tau \, \dot{x}_c(\tau) \frac{d}{d\tau} x(\tau - \tau')
	\end{align}
	
	Using our integration variable $x(\tau - \tau') = x_c(\tau) + \sum_{n=0}^{\infty} c_n \phi_n(\tau)$ and operating strictly with the zero mode $\phi_0(\tau) = C\dot{x}_c(\tau)$, the higher order fluctuations drop out near the classical path: 
	\begin{equation}
		= -C \int_{-T/2}^{T/2} d\tau \, \dot{x}_c(\tau) \dot{x}_c(\tau) + 0 = -S_1^{1/2}
	\end{equation}
	
	This crucially tells us how to replace the problematic integration over $c_0(\tau)$ with a perfectly finite integration over all possible instanton times $\tau'$ from $-T/2$ to $T/2$: 
	\begin{equation}
		\frac{dc_0}{\sqrt{2\pi\hbar}} = \sqrt{\frac{S_1}{2\pi\hbar}} \, d\tau'
	\end{equation}
	
	\section{The Corrected Single Instanton Amplitude}
	Now we can go back and correct all our earlier ``mistakes'' by isolating the zero mode from the fluctuation determinant. 
	Including fluctuations, and after successfully integrating the collective coordinate $\tau'$ over all instanton positions between $-T/2$ and $+T/2$ (which yields a factor of $T$), the correct single instanton amplitude becomes: 
	\begin{equation}
		\mathcal{A}_{n=1} = \int [dx] e^{-\frac{1}{\hbar}(S^{(0)} + S^{(2)})} = \mathcal{N} e^{-\frac{1}{\hbar}S_1} \sqrt{\frac{S_1}{2\pi\hbar}} T \prod_{n=1}^{\infty} \frac{1}{\sqrt{\lambda_n}}
	\end{equation}
	where $\prod_{n=1}^{\infty} \frac{1}{\sqrt{\lambda_n}} \equiv \frac{1}{\sqrt{\det' S''}}$ represents the determinant with the zero mode explicitly omitted. 
	
	To cleanly evaluate this, we define a dimensionless ratio $\Phi$ by normalizing against the standard harmonic oscillator determinant: 
	\begin{equation}
		\Phi \equiv \frac{\det \left[ -\frac{d^2}{d\tau^2} + \omega^2 \right]^{1/2}}{\omega \det' \left[ -\frac{d^2}{d\tau^2} + V''(x_c) \right]^{1/2}}
	\end{equation}
	
	Substituting $\Phi$ into our expression, the single instanton transition amplitude condenses into its final, finite form: 
	\begin{align}
		\mathcal{A}_1 &= \frac{\mathcal{N}}{\det\left[-\frac{d^2}{d\tau^2} + \omega^2\right]^{1/2}} \, \omega \Phi \times \sqrt{\frac{S_1}{2\pi\hbar m}} \, e^{-\frac{S_1}{\hbar}} T \nonumber \\
		&= \left(\frac{\omega}{\pi\hbar}\right)^{1/2} e^{-\frac{\omega T}{2}} \times e^{-\frac{S_1}{\hbar}} K T \nonumber \\
		&= \sqrt{\frac{\omega}{2\pi\hbar \sinh \omega T}} \xrightarrow{T \rightarrow \infty} \left(\frac{\omega}{\pi\hbar}\right)^{1/2} e^{-\frac{\omega T}{2}}
	\end{align}
	where the factor of $T$ arises uniquely from the zero mode integration, a feature that will cleanly exponentiate when we consider the $n > 1$ multi-instanton configurations. 
	
	\section{Dilute Instanton Gas and Energy Splitting}
	For a sufficiently large time interval $T$, the particle can tunnel back and forth many times. 
	In the Dilute Gas Approximation (DGA), we assume these tunneling events are widely separated compared to the core width of a single instanton (which is roughly $\sim \mathcal{O}(1/\omega)$). 
	
	\begin{figure}[h]
		\centering
		\includegraphics[width=0.8\textwidth]{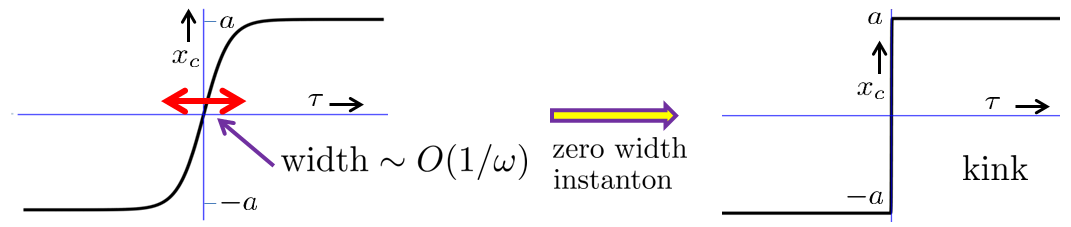}
		\caption{In the Dilute Gas Approximation, the finite-width instanton is effectively treated as a zero-width, instantaneous step function between vacua.}
		\label{fig:zero_width}
	\end{figure}
	
	The path consists of an alternating sequence of $n$ instantons and anti-instantons centered at times $\tau_1 < \tau_2 < \dots < \tau_n$. 
	Since they are assumed to be far apart, the total action is simply the sum of the individual actions: $S_{total} \approx n S_1$. 
	
	\begin{figure}[h]
		\centering
		\includegraphics[width=1\textwidth]{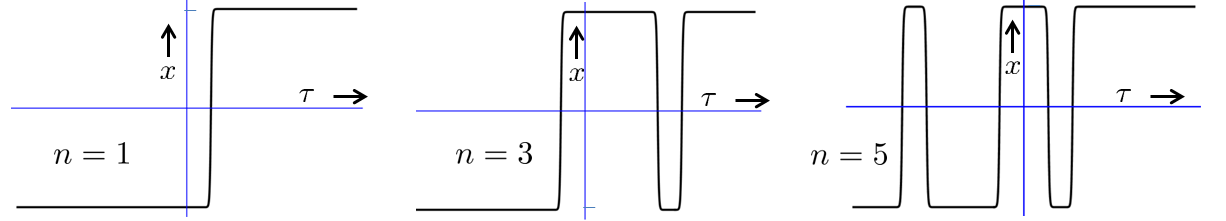}
		\caption{Multi-instanton configurations showing $n=1, n=3,$ and $n=5$ tunneling events. Each kink transitions the system between the degenerate vacua.}
		\label{fig:multi_instanton}
	\end{figure}
	
	The total transition amplitude requires summing over all possible numbers of instantons $n$ and integrating over their non-overlapping positions $\tau_i$: 
	\begin{equation}
		\mathcal{A} \approx \sum_{n=0}^{\infty} \int_{-T/2}^{T/2} d\tau_n \int_{-T/2}^{\tau_n} d\tau_{n-1} \dots \int_{-T/2}^{\tau_2} d\tau_1 \; K^n e^{-\frac{n S_1}{\hbar}}
	\end{equation}
	The time-ordered multiple integral over the instanton positions simply yields the volume of an $n$-dimensional simplex: 
	\begin{equation}
		\int_{-T/2}^{T/2} d\tau_n \int_{-T/2}^{\tau_n} d\tau_{n-1} \dots \int_{-T/2}^{\tau_2} d\tau_1 = \frac{T^n}{n!}
	\end{equation}
	
	Substituting this back into the transition amplitude, we must distinguish between two physical cases based on the number of tunneling events $n$. 
	Because each instanton flips the system from one vacuum to the other, an \textit{even} number of transitions returns the particle to its starting peak, while an \textit{odd} number of transitions leaves the particle at the opposite peak. 
	
	\subsection{Amplitude for Returning to the Same Peak}
	If the particle starts at $-a$ and ends at $-a$ (or starts at $a$ and ends at $a$), $n$ must be an even integer ($n = 0, 2, 4, \dots$). 
	Restricting the sum to even $n$, the amplitude becomes: 
	\begin{equation}
		\mathcal{A}_{\text{same}} = \mathcal{A}_0 \sum_{n \text{ even}} \frac{1}{n!} \left( K T e^{-S_1/\hbar} \right)^n = \mathcal{A}_0 \cosh\left( K T e^{-S_1/\hbar} \right)
	\end{equation}
	where $\mathcal{A}_0$ is the standard harmonic oscillator amplitude for a single well. 
	
	\subsection{Amplitude for Traveling to the Opposite Peak}
	If the particle starts at $-a$ and ends at $a$, $n$ must be an odd integer ($n = 1, 3, 5, \dots$). 
	Restricting the sum to odd $n$, the amplitude is: 
	\begin{equation}
		\mathcal{A}_{\text{opposite}} = \mathcal{A}_0 \sum_{n \text{ odd}} \frac{1}{n!} \left( K T e^{-S_1/\hbar} \right)^n = \mathcal{A}_0 \sinh\left( K T e^{-S_1/\hbar} \right)
	\end{equation}
	
	\subsection{Final Energy Splitting}
	To calculate $K$, we evaluate the primed determinant ratio $\Phi$ using the Gel'fand-Yaglom theorem (detailed below). 
	Multiplying this determinant result by the zero-mode normalization factor $\sqrt{S_1 / 2\pi\hbar}$, we obtain the complete prefactor $K$: 
	\begin{equation}
		K = \omega \sqrt{\frac{S_1}{2\pi\hbar}} \sqrt{12} = \omega \sqrt{\frac{6 S_1}{\pi\hbar}}
	\end{equation}
	
	Using the properties of hyperbolic functions, we can expand our earlier transition amplitudes $\mathcal{A}_{\text{same}}$ and $\mathcal{A}_{\text{opposite}}$ in terms of exponentials: 
	\begin{align}
		\mathcal{A}_{\text{same}} &= \frac{\mathcal{A}_0}{2} \left[ \exp\left(K T e^{-S_1/\hbar}\right) + \exp\left(-K T e^{-S_1/\hbar}\right) \right] \\
		\mathcal{A}_{\text{opposite}} &= \frac{\mathcal{A}_0}{2} \left[ \exp\left(K T e^{-S_1/\hbar}\right) - \exp\left(-K T e^{-S_1/\hbar}\right) \right]
	\end{align}
	
	We compare this to the general spectral decomposition of the imaginary-time propagator at large $T$: $\langle x_f | e^{-HT/\hbar} | x_i \rangle = \sum_m \psi_m(x_f) \psi_m^*(x_i) e^{-E_m T/\hbar}$. 
	Factoring out the unperturbed ground state energy dependence $e^{-E_0 T/\hbar}$ from the single-well amplitude $\mathcal{A}_0$, our transition amplitudes reveal two distinct, closely spaced energy levels: 
	\begin{equation}
		E_{\pm} = E_0 \mp \hbar K e^{-S_1/\hbar}
	\end{equation}
	
	Substituting our exact prefactor $K$, the physical energy splitting between the anti-symmetric first excited state ($E_-$) and the symmetric true ground state ($E_+$) is completely and elegantly resolved: 
	\begin{equation}
		\Delta E = E_- - E_+ = 2\hbar K e^{-S_1/\hbar} = 2\hbar\omega \sqrt{\frac{6 S_1}{\pi\hbar}} e^{-S_1/\hbar}
	\end{equation}
	where $S_1 = \frac{2}{3}m\omega a^2$. 
	This demonstrates how purely non-perturbative instanton tunneling perfectly lifts the classical degeneracy of the symmetric double well. 
	
\begin{figure}[h]
		\centering
		\begin{minipage}{0.48\textwidth}
			\centering
			\includegraphics[width=\linewidth]{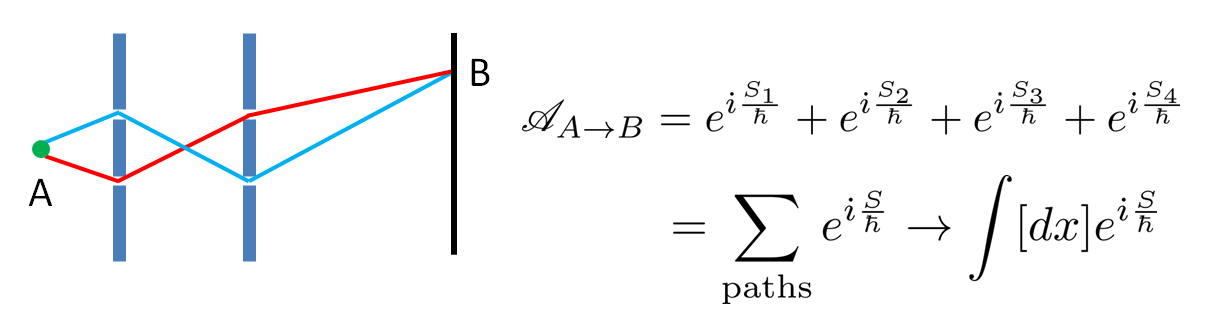}
			\caption{Symmetric eigenstate of the Hamiltonian corresponding to the lower ground state energy $E_+$.}
			\label{fig:symmetric}
		\end{minipage}\hfill
		\begin{minipage}{0.48\textwidth}
			\centering
			\includegraphics[width=\linewidth]{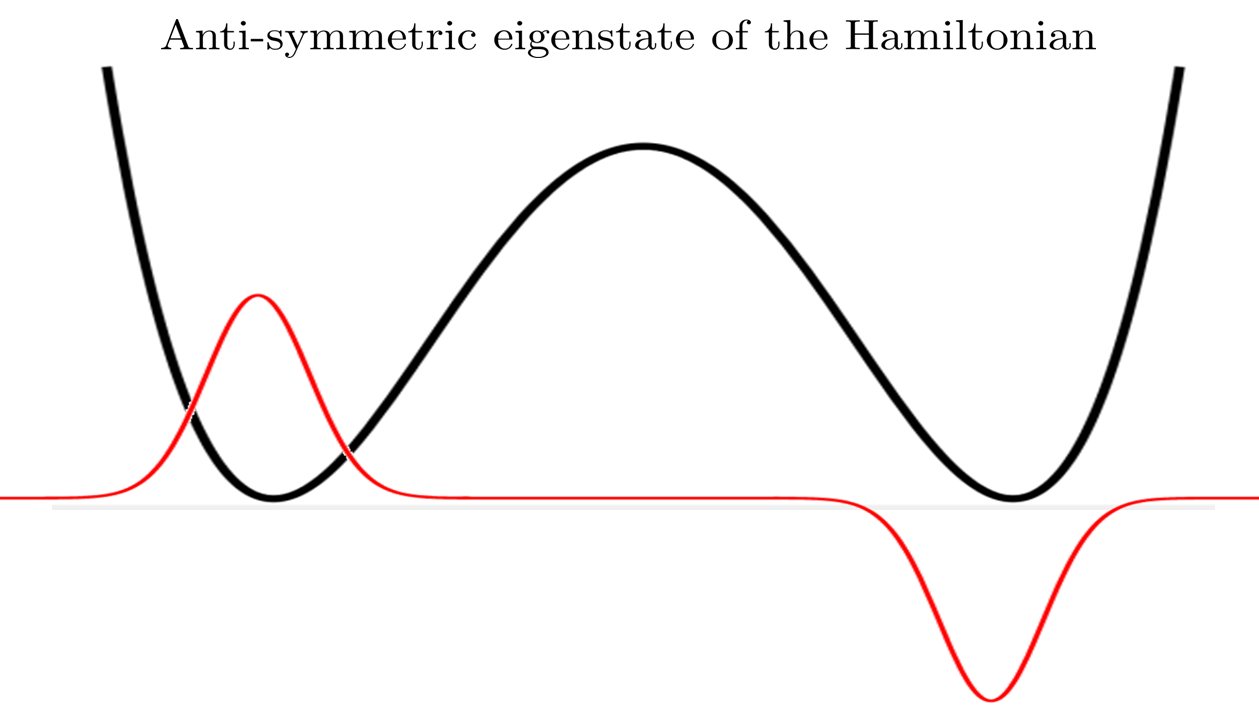}
			\caption{Anti-symmetric eigenstate corresponding to the slightly higher energy $E_-$.}
			\label{fig:antisymmetric}
		\end{minipage}
	\end{figure}

	The lower energy state $E_+$ corresponds to the \textbf{symmetric ground state} (Figure \ref{fig:symmetric}), formed by the positive superposition $|+\rangle \propto |L\rangle + |R\rangle$. 
	The slightly higher energy state $E_-$ corresponds to the \textbf{anti-symmetric first excited state} (Figure \ref{fig:antisymmetric}), formed by the negative superposition $|-\rangle \propto |L\rangle - |R\rangle$. 
	
	\subsection{Validity of the Dilute Gas Approximation}
	The assumption that instantons do not interact is an idealization. 
	In reality, an instanton and an anti-instanton exhibit an attractive interaction that falls off exponentially with their separation distance, $\Delta \tau = \tau_{i+1} - \tau_i$. 
	
	For the DGA to be valid, this interaction must be negligible, which requires the gas to be truly ``dilute.'' 
	This means the average separation between instantons must be exponentially larger than the instanton width: $\langle \Delta \tau \rangle \gg 1/\omega$. 
	
	The density of the instanton gas is proportional to the tunneling rate, governed by the classical action factor $e^{-S_1 / \hbar}$. 
	Therefore, the gas is dilute precisely in the semiclassical limit where the potential barrier is high or wide, leading to: 
	\begin{equation}
		S_1 \gg \hbar
	\end{equation}
	When the barrier is low or narrow, $S_1$ becomes comparable to $\hbar$, the density of the instanton gas increases, and instantons begin to overlap and interact strongly. 
	In this regime, the approximation breaks down (the gas condenses into a ``liquid''), and one must account for multi-instanton interactions or resort to fully non-perturbative techniques. 
	
	\section{Detailed Calculation: The Double-Well Fluctuation Determinant}
	The fluctuation operator for the symmetric double-well instanton takes the form of a Pöschl-Teller potential: 
	\begin{equation}
		\mathcal{O}_c = -\frac{d^2}{d\tau^2} + \omega^2 - \frac{3}{2}\omega^2 \sech^2\left(\frac{\omega}{2}\tau\right)
	\end{equation}
	The determinant of $\mathcal{O}_c$ is evaluated normalized against the free harmonic oscillator reference $\mathcal{O}_v = -\frac{d^2}{d\tau^2} + \omega^2$. 
	By evaluating the operators over a large finite interval $[-T/2, T/2]$, the full determinant ratio can be extracted using the Gel'fand-Yaglom theorem. 
	This allows for the isolation of the asymptotically vanishing zero-mode eigenvalue $\lambda_0$, leading to the exact primed determinant. 
	
	This particular potential belongs to the class of reflectionless potentials, i.e those where every incident wave has perfect transmission $T=1$ independent of its energy. 
	
	\subsection*{Step 1: The Full Determinant Ratio}
	The Gel'fand-Yaglom theorem equates the determinant ratio to the ratio of solutions to the initial value problems $\mathcal{O}_i \phi_i = 0$ with $\phi_i(-T/2) = 0$ and $\dot{\phi}_i(-T/2) = 1$, evaluated at the final boundary $\tau = T/2$. 
	
	For the reference operator $\mathcal{O}_v$, the exact solution is $\phi_v(\tau) = \frac{1}{\omega}\sinh\left(\omega(\tau + T/2)\right)$. 
	Evaluated at the right boundary: 
	\begin{equation}
		\phi_v(T/2) = \frac{\sinh(\omega T)}{\omega} \approx \frac{e^{\omega T}}{2\omega}
	\end{equation}
	
	For the instanton operator $\mathcal{O}_c$, the solution $\phi_c(\tau)$ is constructed using its known unnormalized zero mode, $\psi_0(\tau) = \sech^2(\frac{\omega}{2}\tau)$. 
	The second linearly independent solution is generated via integration, giving: 
	\begin{equation}
		\phi_c(\tau) = \psi_0(-T/2) \psi_0(\tau) \int_{-T/2}^\tau \frac{d\tau'}{\psi_0^2(\tau')}
	\end{equation}
	Evaluating at $\tau = T/2$ and utilizing the even symmetry $\psi_0(-T/2) = \psi_0(T/2) \approx 4e^{-\omega T/2}$: 
	\begin{equation}
		\phi_c(T/2) = \psi_0(T/2)^2 \int_{-T/2}^{T/2} \cosh^4\left(\frac{\omega}{2}\tau'\right) d\tau'
	\end{equation}
	For large $T$, the integrand behaves as $\cosh^4(\frac{\omega}{2}\tau) \approx \frac{1}{16}e^{2\omega\tau}$. 
	Evaluating the integral yields $\approx \frac{e^{\omega T}}{16\omega}$. 
	The boundary solution is therefore strictly independent of $T$: 
	\begin{equation}
		\phi_c(T/2) \approx \left(16 e^{-\omega T}\right) \left( \frac{e^{\omega T}}{16\omega} \right) = \frac{1}{\omega}
	\end{equation}
	The full determinant ratio is exactly: 
	\begin{equation}
		\frac{\det \mathcal{O}_c}{\det \mathcal{O}_v} = \frac{\phi_c(T/2)}{\phi_v(T/2)} = \frac{1/\omega}{e^{\omega T} / 2\omega} = 2e^{-\omega T}
	\end{equation}
	
	\subsection*{Step 2: Isolating the Zero-Mode Eigenvalue}
	The full determinant vanishes exponentially as $T \to \infty$ due to the zero-mode eigenvalue $\lambda_0$. 
	To find $\lambda_0(T)$ on the finite box, the exact zero mode $\psi_0(\tau)$ must be perturbed so it perfectly satisfies the Dirichlet boundaries $\tilde{\psi}_0(\pm T/2) = 0$. 
	
	In the asymptotic tail, $\psi_0(\tau) \approx 4e^{-\omega\tau}$. 
	To force the function to zero at $T/2$, a complementary free solution $\propto e^{\omega\tau}$ is subtracted: 
	\begin{equation}
		\tilde{\psi}_0(\tau) \approx 4e^{-\omega\tau} - 4e^{-\omega T}e^{\omega\tau}
	\end{equation}
	The standard formula for the lowest eigenvalue derived via integration by parts is: 
	\begin{equation}
		\lambda_0 \approx \frac{2 \psi_0(T/2) |\tilde{\psi}_0'(T/2)|}{\int_{-\infty}^\infty \psi_0^2(\tau) d\tau}
	\end{equation}
	Evaluating the boundary derivative gives $\vert{}\tilde{\psi}_0'(T/2)\vert{} = 8\omega e^{-\omega T/2}$. 
	The normalization integral evaluates to $\int_{-\infty}^\infty \sech^4\left(\frac{\omega}{2}\tau\right) d\tau = \frac{8}{3\omega}$. 
	Substituting these components gives the asymptotic eigenvalue: 
	\begin{equation}
		\lambda_0 = \frac{2 \left(4e^{-\omega T/2}\right) \left(8\omega e^{-\omega T/2}\right)}{8 / 3\omega} = 24\omega^2 e^{-\omega T}
	\end{equation}
	
	\subsection*{Step 3: The Primed Determinant}
	The primed determinant explicitly omits the zero mode, $\det \mathcal{O}_c = \lambda_0 \det' \mathcal{O}_c$. 
	Dividing the full determinant ratio by $\lambda_0$ isolates the finite contribution: 
	\begin{equation}
		\frac{\det' \mathcal{O}_c}{\det \mathcal{O}_v} = \frac{2e^{-\omega T}}{24\omega^2 e^{-\omega T}} = \frac{1}{12\omega^2}
	\end{equation}
	Raising this ratio to the inverse square root yields: 
	\begin{equation}
		\left[ \frac{\det' \mathcal{O}_c}{\det \mathcal{O}_v} \right]^{-1/2} = \sqrt{12\omega^2} = \omega\sqrt{12}
	\end{equation}
	Working in dimensionless units where $\omega = 1$ (or extracting the natural frequency into the ratio definition $\Phi$), this produces the exact statement shown in the text: 
	\begin{equation}
		\left[ \frac{\det' \mathcal{O}_c}{\det \mathcal{O}_v} \right]^{-1/2} = \sqrt{12} = 2\sqrt{3}
	\end{equation}

	\newpage
	\part{Decay of a Metastable State (False Vacuum)}
	
	\section{Reminder from Basic Quantum Mechanics}
	We begin with a fundamental reminder from basic quantum mechanics. 
	The Hamiltonian $\hat{H}$ is a hermitian operator if $\hat{H} = \hat{H}^\dagger$. 
	Therefore, its eigenvalues are strictly real, and physical states inherently preserve their norm over time. 
	
	We can explicitly check this time evolution: 
	\begin{equation}
		|\psi(t)\rangle = e^{-\frac{i}{\hbar}\hat{H}t}|\psi(0)\rangle
	\end{equation}
	Taking the inner product yields: 
	\begin{equation}
		\therefore \langle \psi(t)|\psi(t) \rangle = \langle \psi(0)|e^{\frac{i}{\hbar}(\hat{H}^\dagger - \hat{H})t}|\psi(0) \rangle = \langle \psi(0)|\psi(0) \rangle
	\end{equation}
	
	But even remarkably simple Hamiltonians, such as the free particle kinetic energy $\frac{\hat{p}^2}{2m}$, can fail to be hermitian under certain conditions! 
	Let's see why by examining the integration by parts of the expectation value: 
	\begin{align}
		\frac{1}{2m} \int dx (\hat{p}^2 \psi^*)\psi &= -\frac{\hbar^2}{2m} \int dx \left(\frac{d^2}{dx^2}\psi^*\right)\psi \nonumber \\
		&= -\frac{\hbar^2}{2m} \int dx \frac{d}{dx} \left\{ \frac{d\psi^*}{dx}\psi - \psi^* \frac{d\psi}{dx} \right\} - \frac{\hbar^2}{2m} \int dx \, \psi^* \frac{d^2}{dx^2}\psi
	\end{align}
	The surface term is normally zero, rendering the operator hermitian: 
	\begin{equation}
		0 = \frac{1}{2m} \int dx \, \psi^* \hat{p}^2 \psi
	\end{equation}
	However, this is true only if the boundary terms vanish, i.e., no probability current leaks out of the system. 
	
	If current does leak out (as in a tunneling escape), the Hamiltonian ceases to be purely hermitian. 
	Else, the energy acquires an imaginary component, $E = \epsilon_R - i\Gamma/2$, and the wavefunction becomes a decaying state: 
	\begin{equation}
		\psi(t) = \psi(0) e^{-\frac{i}{\hbar}\epsilon_R t} e^{-\frac{\Gamma}{2\hbar}t}
	\end{equation}
	
	\section{Decay of a Metastable State}
	We now consider the decay of a metastable state, often referred to as a ``false vacuum''. 
	Unlike the symmetric double well, the potential $V(x)$ here has a local minimum (Vacuum A) that is energetically higher than a global minimum (Vacuum B). 
	
	\begin{figure}[h]
		\centering
		\includegraphics[width=0.9\textwidth]{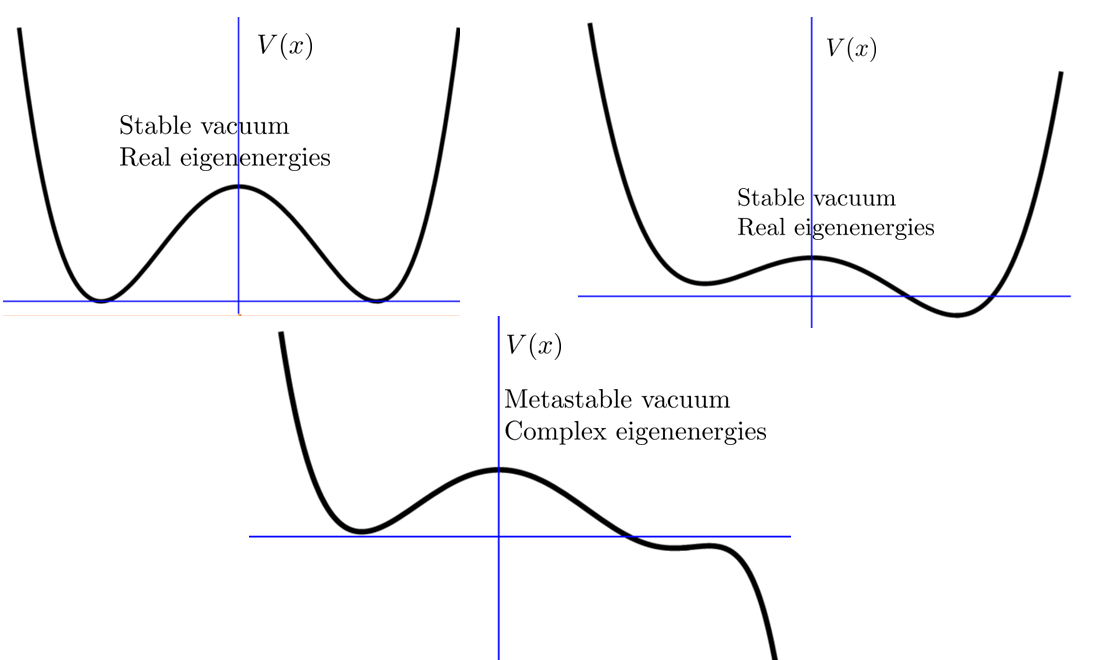}
		\caption{A potential featuring a stable vacuum and a metastable vacuum. States localized in the metastable vacuum have complex eigenenergies due to the possibility of tunneling decay.}
		\label{fig:metastable}
	\end{figure}
	
	To calculate the decay rate $\Gamma$, we utilize the Euclidean path integral. 
	We Wick rotate to imaginary time, which inverts the potential to $-V(x)$. 
	The metastable vacuum becomes a local hilltop, while the true vacuum becomes a deep valley. 
	
	\begin{figure}[h]
		\centering
		\includegraphics[width=0.7\textwidth]{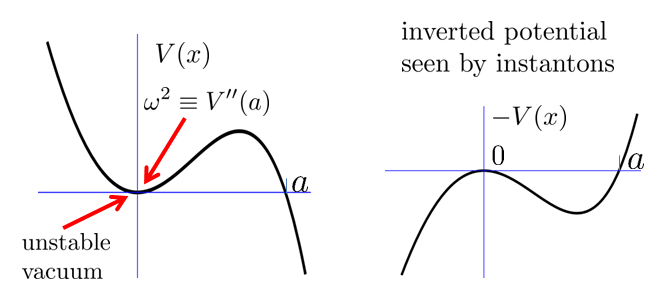}
		\caption{The inverted potential $-V(x)$ seen by instantons in imaginary time.}
		\label{fig:inverted_metastable}
	\end{figure}
	
	The relevant classical trajectory for this geometry is the ``bounce''. 
	The instanton travels from the false vacuum at $x=0$ to the classical turning point $x=a$ and reflects back. 
	\begin{itemize}
		\item The departure time from $0$ is $\tau = -T/2$. 
		\item The arrival time at the turning point $a$ is $\tau = 0$. 
		\item The instanton returns to $x=0$ at $\tau = T/2$. 
	\end{itemize}
	
	\begin{figure}[h]
		\centering
		\includegraphics[width=0.7\textwidth]{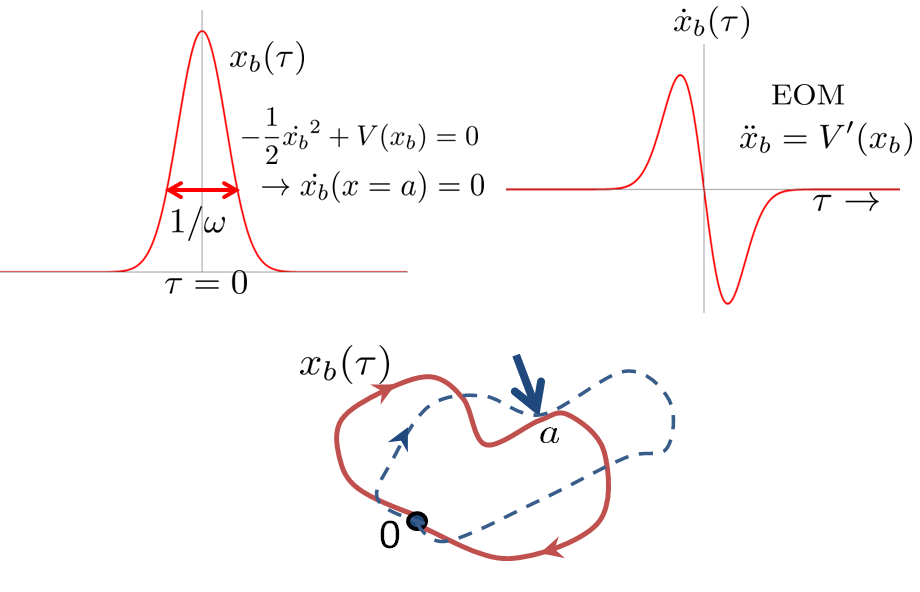}
		\caption{The bounce trajectory $x_b(\tau)$. Notice that it starts at 0, reaches $a$, and returns to 0.}
		\label{fig:bounce_profile}
	\end{figure}
	
	Conservation of zero energy in imaginary time gives: 
	\begin{equation}
		-\frac{1}{2}\dot{x}_b^2 + V(x_b) = 0 \implies \dot{x}_b(x=a) = 0 \quad \text{at} \quad \tau = 0
	\end{equation}
	The equation of motion (EOM) remains: $\ddot{x}_b = V'(x_b)$. 
	
	\section{The Fluctuation Determinant and the Negative Mode}	
	Formally, the expanded action is exactly as it was in lecture \#2: 
	\begin{equation}
		S_{total} = S_{classical}^{(0)} + S_{fluctuation}^{(2)}
	\end{equation}
	\begin{equation}
		S_{fluctuation}^{(2)} = \frac{1}{2} \int_{\tau_a}^{\tau_b} d\tau \, \xi(\tau) \left( -\frac{d^2}{d\tau^2} + V''(x_b) \right) \xi(\tau)
	\end{equation}
	where the classical bounce action is: 
	\begin{equation}
		S_{classical}^{(0)} \equiv S_b^{(0)} = \int_{-T/2}^{T/2} d\tau \left( \frac{1}{2}\dot{x}_b^2 + V(x_b) \right)
	\end{equation}
	subject to the boundary conditions $x_b(-T/2) = x_b(T/2) = 0$, and the turning point $x_b(0) = a$. 
	
	The bounce velocity $\dot{x}_b(\tau)$, by virtue of being a derivative of the solution of the EOM, must be an eigenfunction of the fluctuation operator $S''$ with a zero eigenvalue. 
	The proof follows exactly as before: 
	\begin{equation}
		\left( -\frac{d^2}{d\tau^2} + V''(x_b) \right) \frac{dx_b}{d\tau} = \frac{d}{d\tau} \left( -\frac{d^2 x_b}{d\tau^2} + V'(x_b) \right) = 0
	\end{equation}
	Note that $\dot{x}_b(\tau)$ gracefully obeys the correct boundary conditions, i.e., $\dot{x}_b(-T/2) = \dot{x}_b(+T/2) = 0$. 
	
	But now note a crucial difference: while $\dot{x}_b(\tau)$ has a zero eigenvalue, it also has a node at $\tau=0$ because the velocity goes to zero when the bounce turns around at $x=a$. 
	According to Sturm-Liouville theory, the ground state eigenfunction of a one-dimensional operator has no nodes. 
	Because the zero mode has one node, there must exist some still lower eigenfunction that is without a node, and therefore it must possess a strictly negative eigenvalue ($\lambda_{-1} < 0$). 
	
	\section{Complex Variables and the Saddle Point}
	
	\begin{figure}[h]
		\centering
		\includegraphics[width=0.8\textwidth]{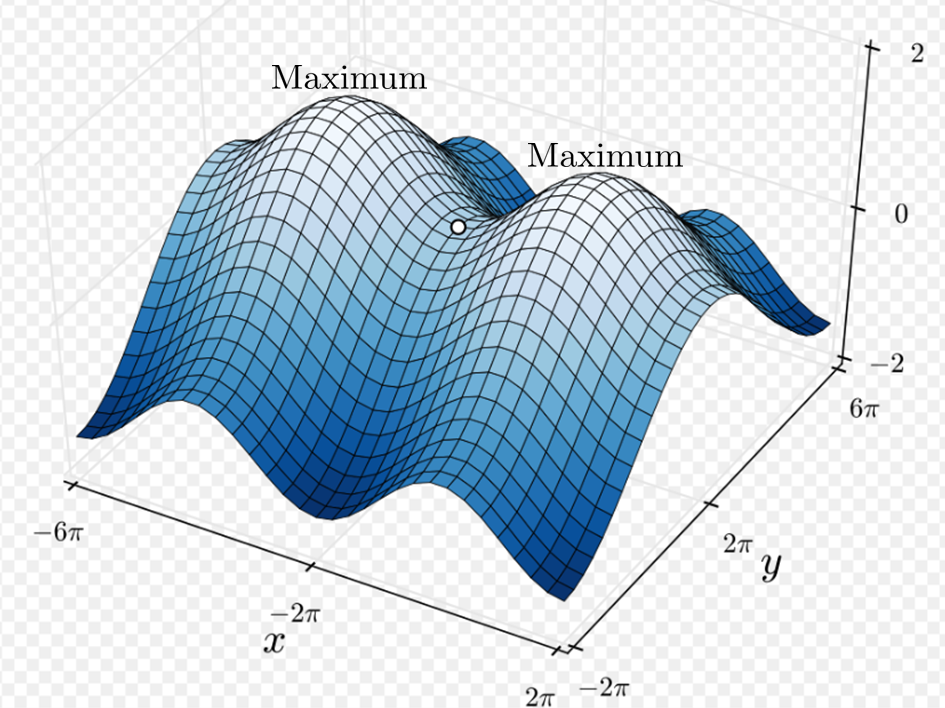}
		\caption{A generic multi-dimensional surface showing local maxima, minima, and saddle points.}
		\label{fig:surface}
	\end{figure}
	
	The existence of a negative eigenvalue fundamentally changes the path integral. 
	When expanding the fluctuations in normal modes, the integral over the coefficient $c_{-1}$ associated with the negative eigenvalue $\lambda_{-1}$ becomes: 
	\begin{equation}
		\int_{-\infty}^{\infty} d c_{-1} \exp\left( - \frac{\lambda_{-1}}{2\hbar} c_{-1}^2 \right)
	\end{equation}
	Because $\lambda_{-1}$ is negative, the exponent is positive, and this Gaussian integral blows up to infinity along the real axis. 
	
	To resolve this, we must briefly review complex variables and integration. 
	The classical bounce represents a saddle point of the Euclidean action in the infinite-dimensional configuration space (similar to the geometry shown in Figure \ref{fig:surface}). 
	To define the integral, we must analytically continue the integration contour by applying the method of steepest descent. 
	
	\begin{figure}[h]
		\centering
		\includegraphics[width=0.8\textwidth]{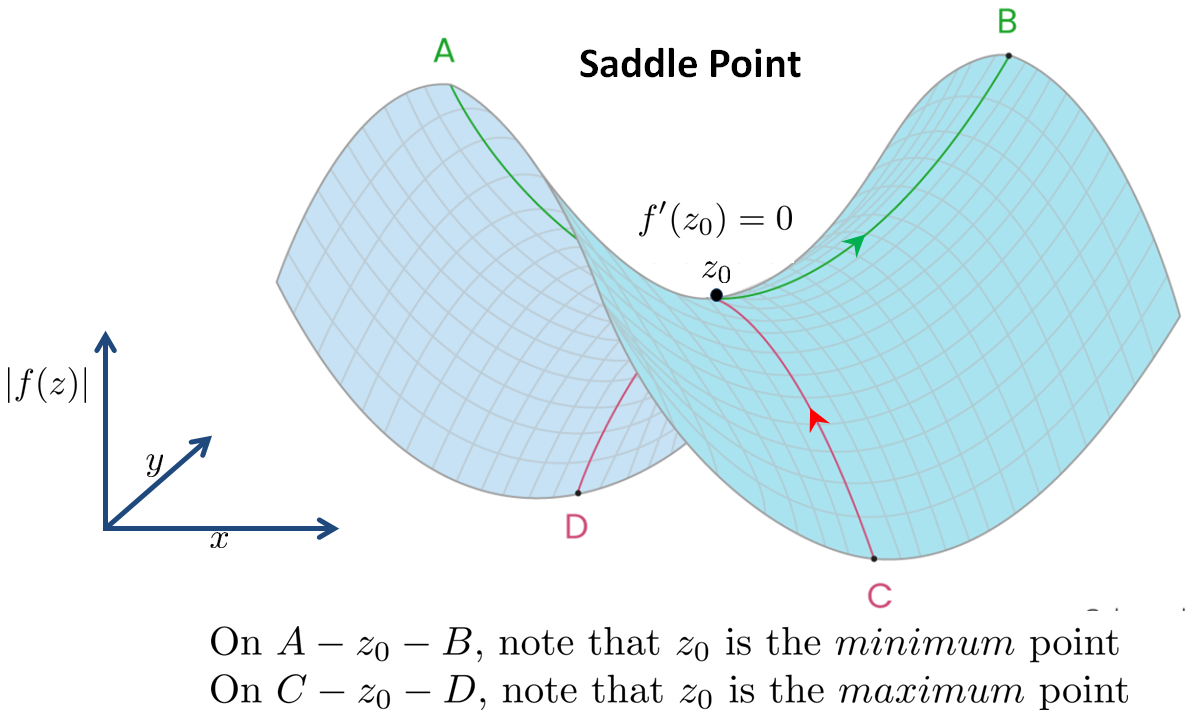}
		\caption{A saddle point $x_0$ where $f'(x_0) = 0$. The path of steepest descent passes through the saddle point, transforming the divergent integration into a convergent Gaussian.}
		\label{fig:saddle}
	\end{figure}
	
	Look at the space of all paths satisfying $x(-T/2) = x(T/2) = 0$ on which $x(\tau)$ takes its maximum at $\tau=0$. 
	Let $c \equiv x(\tau=0)$. 
	We will parameterize paths by their value of $c$. 
	
	A definite value of action is associated with each path, i.e., each value of $c$. 
	The path which just reaches $a$ extremizes $S(c)$. 
	\begin{equation}
		\left. \frac{dS(c)}{dc} \right|_{c=a} = 0
	\end{equation}
	
	To give meaning to $\int dc$ we shall have to continue a real integral into the complex plane, i.e. use analytic continuation as depicted in Figure \ref{fig:saddle}. 
	
	Loss of probability is associated with $c > a$. 
	We should therefore expect an imaginary part in the probability amplitude. 
	
	\begin{figure}[h]
		\centering
		\includegraphics[width=0.5\textwidth]{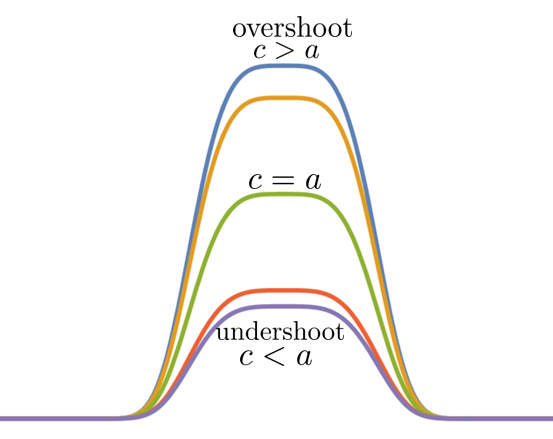}
		\caption{Paths characterized by their maximum amplitude $c$. Undershoot paths have $c < a$, the classical bounce path has $c=a$, and overshoot paths have $c>a$. The overshoot paths represent a loss of probability, indicating an imaginary component to the energy.}
		\label{fig:overshoot}
	\end{figure}
	
	A direct application of Cauchy's integral theorem to deform the integration contour from the real axis to the imaginary axis actually fails because the bounding arcs at infinity exponentially diverge. To rigorously derive the imaginary prefactor, we must explicitly evaluate the contour sections to see why direct deformation breaks down, and then apply Cauchy's theorem to the analytic continuation of the potential parameters.
	
	\textbf{1. The Failure of Direct Contour Deformation} \\
	Let the negative mode integral be parameterized by $u = c_{-1}$. We are attempting to evaluate $I = \int_{-\infty}^\infty du \, e^{\frac{1}{2\hbar}|\lambda_{-1}| u^2}$. Consider a closed Cauchy contour $\Gamma$ in the first quadrant consisting of three segments:
	\begin{itemize}
		\item \textbf{$C_1$ (Real Axis):} $u$ from $0$ to $R$. The integral is $\int_0^R dx \, e^{\frac{1}{2\hbar}|\lambda_{-1}|x^2}$. As $R \to \infty$, this integral diverges exponentially.
		\item \textbf{$C_2$ (Arc at Infinity):} $u = R e^{i\theta}$ from $\theta = 0$ to $\pi/2$. The exponent becomes $\frac{1}{2\hbar}|\lambda_{-1}| R^2 (\cos 2\theta + i\sin 2\theta)$. In the sector $0 \le \theta < \pi/4$, $\cos 2\theta > 0$, so the integrand grows exponentially to infinity.
		\item \textbf{$C_3$ (Imaginary Axis):} $u = iy$ from $R$ down to $0$. The integral is $\int_R^0 (i\,dy) e^{-\frac{1}{2\hbar}|\lambda_{-1}|y^2}$. As $R \to \infty$, this converges perfectly to $-i \sqrt{\frac{\pi\hbar}{2|\lambda_{-1}|}}$.
	\end{itemize}
	Cauchy's theorem dictates $\oint_\Gamma = \int_{C_1} + \int_{C_2} + \int_{C_3} = 0$. Because $C_1$ and $C_2$ independently diverge to infinity, we cannot algebraically isolate $C_1$ to equal $-C_3$. Therefore, the real-axis integral is strictly undefined, and we cannot simply "deform" it into the imaginary axis.
	
	\textbf{2. Rigorous Proof via Analytic Continuation} \\
	The mathematically rigorous resolution, formulated by Langer and utilized by Callan and Coleman, defines the false vacuum decay rate by analytically continuing the integral from a strictly stable state. 
	
	\begin{enumerate}
		\item Assume a stable potential where the lowest eigenvalue is positive, $\lambda > 0$. The path integral over the mode amplitude $u$ is perfectly convergent along the real axis: $I(\lambda) = \int_{-\infty}^\infty du \exp\left(-\frac{\lambda}{2\hbar} u^2\right) = \sqrt{\frac{2\pi\hbar}{\lambda}}$.
		\item We analytically continue the parameter $\lambda$ from positive to negative (representing the metastable state) via a continuous path in the complex plane: $\lambda = |\lambda| e^{-i\theta}$ from $\theta = 0$ to $\theta = \pi$.
		\item We apply Cauchy's theorem dynamically. To maintain the integral's convergence at infinity as $\theta$ rotates, we must simultaneously rotate the integration contour $u = r e^{i\alpha}$ such that the real part of the exponent remains strictly negative. 
		\item The exponent is $-\frac{|\lambda|}{2\hbar} r^2 e^{i(2\alpha - \theta)}$. For the integral to remain bounded and convergent, we must enforce $2\alpha - \theta = 0$, meaning the contour must rotate by $\alpha = \theta/2$.
		\item As the physical parameter reaches the negative mode ($\theta = \pi$), the integration contour is forced to rotate to $\alpha = \pi/2$. The convergent steepest descent contour $C_{sd}$ is exactly the imaginary axis ($u = i y$).
		\item Evaluating the integral along this convergent contour yields:
		\begin{equation}
			I = \int_{-\infty}^\infty (i\,dy) \exp\left(-\frac{|\lambda_{-1}|}{2\hbar} y^2\right) = i \sqrt{\frac{2\pi\hbar}{|\lambda_{-1}|}} 
		\end{equation}
		This introduces the factor of $i$ that defines the imaginary energy component and the resulting decay rate $\Gamma$.
	\end{enumerate}
	
	\begin{figure}[h]
		\centering
		\begin{tikzpicture}[scale=1.5, >=stealth]
			
			\fill[red!15] (-3,-3) -- (3,3) -- (3,-3) -- cycle;
			\fill[red!15] (-3,-3) -- (-3,3) -- (3,3) -- cycle; 
			\fill[blue!15] (0,0) -- (3,3) -- (-3,3) -- cycle;
			\fill[blue!15] (0,0) -- (3,-3) -- (-3,-3) -- cycle;
			
			\draw[->, thick, black!70] (-3.5,0) -- (3.5,0) node[right, text=black] {$\text{Re}(u)$};
			\draw[->, thick, black!70] (0,-3.5) -- (0,3.5) node[above, text=black] {$\text{Im}(u)$};
			
			\draw[dashed, gray, thick] (-3,-3) -- (3,3);
			\draw[dashed, gray, thick] (-3,3) -- (3,-3);
			
			\node[red!80!black, font=\bfseries] at (2.5, 0.4) {DIVERGENT};
			\node[red!80!black, font=\bfseries] at (-2.5, 0.4) {DIVERGENT};
			\node[blue!80!black, font=\bfseries] at (0.8, 2.5) {CONVERGENT};
			\node[blue!80!black, font=\bfseries] at (0.8, -2.5) {CONVERGENT};
			
			\draw[->, very thick, red!80!black] (-3.2,0) -- (3.2,0);
			
			\draw[->, very thick, blue!80!black] (0,-3.2) -- (0,3.2) node[above right] {$u = iy$};
			
			\draw[->, ultra thick, purple] (1.5, 0) arc (0:90:1.5);
			\node[purple, font=\bfseries] at (1.8, 1.8) {$\alpha = \theta/2 = \pi/2$};
			
			\filldraw[black] (0,0) circle (2pt) node[below right, font=\bfseries] {Saddle Point $x_0$};
			
		\end{tikzpicture}
		\caption{The complex $u$-plane for the negative mode integrand $e^{\frac{1}{2\hbar}|\lambda_{-1}| u^2}$. The real axis lies entirely in the divergent sector, preventing direct Cauchy closure via arcs at infinity. Analytic continuation of the parameter space forces the contour to rotate by $\pi/2$ into the convergent steepest-descent valley along the imaginary axis.}
		\label{fig:contour_deformation_rigorous}
	\end{figure}
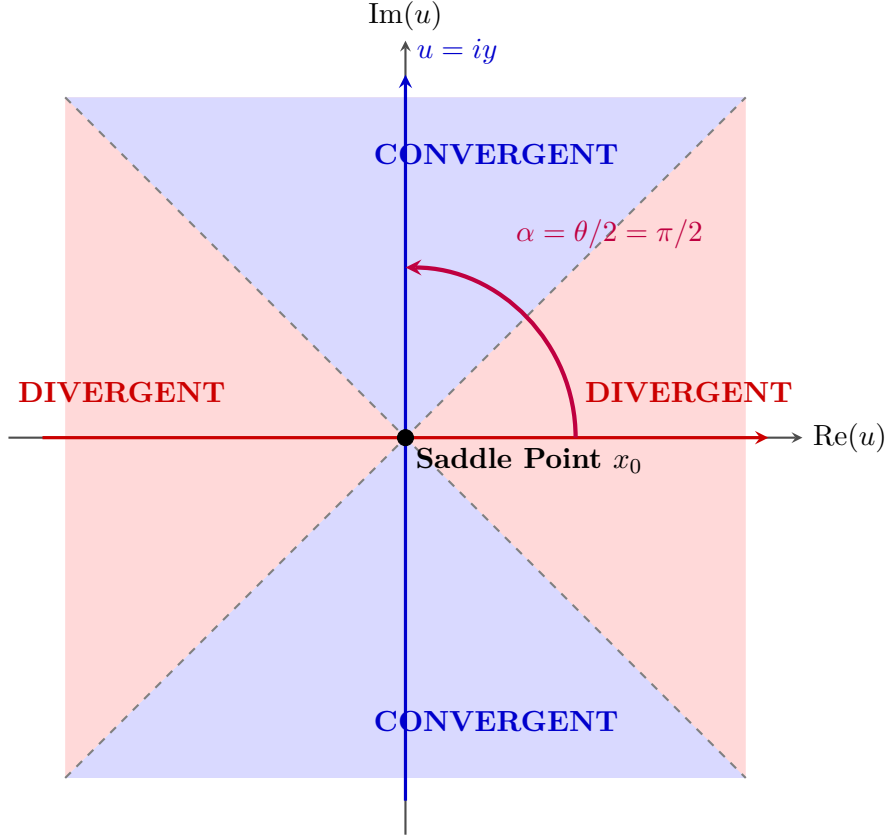
	
	\section{The Dilute Bounce Gas and Comparison with Stable States}
	
	In our previous analysis of the symmetric double well (Part III), we found that to obtain the full physical picture over a large time interval $T$, we had to sum over configurations containing an arbitrary number of widely separated instantons and anti-instantons. 
	We must apply a similar Dilute Gas Approximation (DGA) here for the metastable state, but the mathematical behavior and physical implications are profoundly different. 
	
	\subsection{Summing the Multi-Bounce Configurations}
	For the false vacuum, the relevant classical configuration is the bounce. 
	Over a long time interval $T$, the system can undergo multiple independent, widely separated bounce events. 
	Let us consider a configuration with $n$ such bounces. 
	Because the bounces are assumed to be widely separated, the total action is approximately $n S_b$. 
	As before, integrating over the independent temporal locations of the $n$ bounces yields a volume phase-space factor of $T^n / n!$. 
	
	However, there is a fundamental difference in the fluctuation determinant: \textit{each} individual bounce configuration possesses a single negative mode. 
	Therefore, the analytic continuation of the fluctuation determinant ($c_{-1} \to i c_{-1}$) contributes a factor of $i$ for every single bounce in the sequence. 
	
	If we define the magnitude of the single-bounce prefactor (incorporating the zero mode Jacobian and the absolute value of the determinant ratio) as $K$, the contribution of $n$ bounces to the total transition amplitude is: 
	\begin{equation}
		\mathcal{A}_n = \mathcal{A}_0 \frac{1}{n!} \left( i K T e^{-S_b/\hbar} \right)^n
	\end{equation}
	where $\mathcal{A}_0$ is the amplitude for the trivial background (no bounces). 
	
	Summing over all possible numbers of bounces $n$ from $0$ to $\infty$ exponentiates this contribution perfectly: 
	\begin{equation}
		\mathcal{A}_{total} = \mathcal{A}_0 \sum_{n=0}^{\infty} \frac{1}{n!} \left( i K T e^{-S_b/\hbar} \right)^n = \mathcal{A}_0 \exp\left( i K T e^{-S_b/\hbar} \right)
	\end{equation}
	
	\subsection{Comparison: Decay Rate vs. Energy Splitting}
	We can now directly compare this result with the standard spectral decomposition, $\mathcal{A} \propto e^{-E T/\hbar}$, where the energy is shifted from the harmonic ground state by $E = E_0 + \Delta E$. 
	
	Comparing the exponents, we identify the energy shift $\Delta E$ for the false vacuum: 
	\begin{equation}
		-\frac{\Delta E \, T}{\hbar} = i K T e^{-S_b/\hbar} \implies \Delta E = -i \hbar K e^{-S_b/\hbar}
	\end{equation}
	Notice that the energy shift is purely imaginary! 
	Recalling from Section 1 that a complex energy $E = \epsilon_R - i\Gamma/2$ indicates a decaying state, we can immediately identify the decay rate: 
	\begin{equation}
		\Gamma = -2 \, \text{Im}(\Delta E) = 2 \hbar K e^{-S_b/\hbar}
	\end{equation}
	Expanding $K$ explicitly, we arrive at the famous ''Grand Result'' for the decay of a false vacuum: 
	\begin{equation}
		\Gamma = \left| \frac{\det' \left( -\partial_\tau^2 + V''(x_b) \right)}{\det \left( -\partial_\tau^2 + \omega^2 \right)} \right|^{-1/2} e^{-S_b/\hbar}
	\end{equation}
	
	\textbf{Contrast this with the stable degenerate vacua (the symmetric double well):} 
	\begin{itemize}
		\item \textbf{Spectrum of Modes:} In the double well, single instantons and anti-instantons represent monotonic transitions between vacua. Their velocity profiles do not cross zero, so they have no nodes. Consequently, they possess no negative modes. 
		\item \textbf{Nature of the Summation:} Because there are no negative modes, the individual contributions to the path integral in the double well are strictly real. Summing over even and odd numbers of instantons yields real $\cosh$ and $\sinh$ functions, rather than a complex exponential phase. 
		\item \textbf{Calculational Goal:} In the double well, the real sum translates to a strictly \textit{real} energy shift $\pm \Delta E$. The primary calculational goal is finding the \textbf{energy splitting} that lifts the classical degeneracy to form stable symmetric and anti-symmetric ground states. 
		\item \textbf{Physical Interpretation:} In the metastable false vacuum, there is no second degenerate well to safely transition into and remain; the bounce represents a temporary tunneling excursion under the barrier and a subsequent materialization into a continuum or unbound state. The negative mode of the bounce drives the sum into a complex phase, shifting the calculational goal from finding a stable energy splitting to finding the \textbf{decay rate} ($\Gamma$) of an unstable state. 
	\end{itemize}
	
	\section{Detailed Calculation: A Toy Model with a Cubic Potential}
	The most famous and mathematically elegant toy model for exactly calculating the fluctuation determinant of a false vacuum decay is the cubic potential in one-dimensional quantum mechanics. 
	Because the resulting fluctuation operator takes the form of the exactly solvable Pöschl-Teller potential, the entire calculation---including the negative mode, the zero mode, and the exact determinant ratio---can be carried out analytically using the Gel'fand-Yaglom theorem. 
	The explicit formulation and exact solution of this toy model was originally derived by Callan and Coleman. 
	
	\subsection{The Cubic Potential and the Exact Bounce}
	We define a potential with a metastable minimum at $x=0$ and a cubic barrier that drops to negative infinity, normalized as: 
	\begin{equation}
		V(x) = \frac{1}{2}m\omega^2 x^2 - \frac{\lambda}{3} x^3
	\end{equation}
	In imaginary time ($\tau = it$), the potential is inverted ($-V(x)$), turning the local minimum into a hilltop and the barrier into a valley. 
	
	The classical equation of motion in imaginary time is $m\ddot{x} = V'(x)$. 
	For the cubic potential, this yields a closed-form analytic bounce trajectory that starts at $x=0$, reaches the turning point, and returns to $x=0$: 
	\begin{equation}
		x_b(\tau) = \frac{3m\omega^2}{2\lambda} \text{sech}^2\left( \frac{\omega}{2} \tau \right)
	\end{equation}
	This trajectory describes a localized ``lump'' or ``bounce'' in imaginary time. 
	The classical Euclidean action for this bounce evaluates exactly to $S_b = \frac{3m^3\omega^5}{5\lambda^2}$. 
	
	\subsection{The Pöschl-Teller Fluctuation Operator}
	To find the decay rate, we must evaluate the fluctuation determinant. 
	We expand the action to second order around the bounce $x_b(\tau)$. 
	The fluctuation operator is: 
	\begin{equation}
		\mathcal{O}_c = -m\frac{d^2}{d\tau^2} + V''(x_b(\tau))
	\end{equation}
	By taking the second derivative of the cubic potential and substituting the bounce solution, the operator becomes: 
	\begin{equation}
		\mathcal{O}_c = -m\frac{d^2}{d\tau^2} + m\omega^2 \left[ 1 - \frac{3}{\cosh^2\left(\frac{\omega}{2}\tau\right)} \right]
	\end{equation}
	This is the famous Pöschl-Teller operator. 
	It is one of the few quantum mechanical potentials whose entire spectrum (bound states and scattering states) is known exactly. 
	
	\subsection{The Exact Spectrum}
	Because the operator is analytically solvable, we can immediately identify its discrete spectrum, which perfectly matches the theoretical requirements for false vacuum decay: 
	\begin{itemize}
		\item \textbf{The Negative Mode ($n=0$):} There is exactly one state with a negative eigenvalue ($\lambda_0 = -\frac{5}{4}m\omega^2$). 
		This state has no nodes, and its analytic continuation is responsible for the imaginary part of the energy (the decay rate). 
		\item \textbf{The Zero Mode ($n=1$):} There is exactly one state with a zero eigenvalue ($\lambda_1 = 0$). 
		This is proportional to the derivative of the bounce, $\dot{x}_b(\tau)$, which has exactly one node. 
		It corresponds to time-translation invariance. 
		\item \textbf{The Positive Mode ($n=2$):} There is a single positive bound state with eigenvalue $\lambda_2 = \frac{3}{4}m\omega^2$. 
		\item \textbf{The Continuum:} A continuous spectrum of positive eigenvalues begins at $\lambda = m\omega^2$, corresponding to stable fluctuations. 
	\end{itemize}
	
	\subsection{Applying Gel'fand-Yaglom to Find $K$}
	Because we know the exact differential equation for the Pöschl-Teller potential, we do not need to multiply infinite series of eigenvalues. 
	We can apply the Gel'fand-Yaglom theorem to find the ratio of the determinant of $\mathcal{O}_c$ (the bounce) to $\mathcal{O}_{SHO} = -m\frac{d^2}{d\tau^2} + m\omega^2$ (the free harmonic oscillator). 
	
	The theorem maps the determinant ratio to the asymptotic behavior of the zero mode $\psi_0(\tau) \propto \dot{x}_b(\tau)$. 
	Properly normalized such that $\int \psi_0^2 d\tau = 1$, the zero mode has the asymptotic form $\psi_0(\tau) \to C e^{-\omega \tau}$ as $\tau \to \infty$. 
	
	Using the initial value formulation of the Gel'fand-Yaglom theorem, the exact, finite determinant ratio (with the zero mode explicitly removed) evaluates to: 
	\begin{equation}
		\left| \frac{\det' \mathcal{O}_c}{\det \mathcal{O}_{SHO}} \right| = \frac{1}{60\omega^2}
	\end{equation}
	
	Substituting this into our grand result for the decay rate prefactor $K$: 
	\begin{equation}
		K = \omega \left| \frac{\det' \mathcal{O}_c}{\omega^{-2} \det \mathcal{O}_v} \right|^{-1/2} \sqrt{\frac{S_b}{2\pi\hbar}} = \omega \sqrt{60} \sqrt{\frac{S_b}{2\pi\hbar}} = \omega \sqrt{\frac{30 S_b}{\pi\hbar}}
	\end{equation}
	This exact calculation perfectly resolves the macroscopic quantum tunneling out of a metastable well, yielding a final decay rate of $\Gamma = \frac{\omega}{2\pi} \sqrt{\frac{30 S_b}{\pi\hbar}} e^{-S_b/\hbar}$ where $S_b = \frac{3m^3\omega^5}{5\lambda^2}$.

	\newpage
	\part{Tunneling in Quantum Field Theory}
	
	\section{Introduction: Tunneling in QFT}
	We now extend our tunneling analysis to Quantum Field Theory. 
	This involves systems with an infinite number of degrees of freedom, making the physics much richer. 
	It is important to emphasize at the outset that this lecture is fundamentally an exposition of the seminal work by Sidney Coleman. 
	His masterly formulation of false vacuum decay, along with his brilliantly intuitive mechanical analogies, forms the rigorous bedrock of our current understanding. 
	
	Examples of macroscopic tunneling and instanton effects span various domains of physics, including the stability of the electroweak vacuum, spin glasses, chemical reactions, multiple vacua in QCD, and superfluid $^3$He. 
	
	\subsection{Qualitative Example: Boiling and Nucleation}
	A familiar macroscopic analogy for vacuum decay is the boiling of water at $T \geq 100^\circ\text{C}$. 
	When water is heated, it does not instantly transition into vapor everywhere simultaneously. 
	Instead, localized bubbles of steam form through a process called \textit{nucleation}. 
	
	Inside the fluid, a bubble experiences competing forces: the vapor pressure inside the bubble pushes outward, attempting to expand it, while the ambient pressure of the fluid and the surface tension of the bubble wall pull inward, attempting to collapse it. 
	
	\begin{figure}[h]
		\centering
		\includegraphics[width=0.6\textwidth]{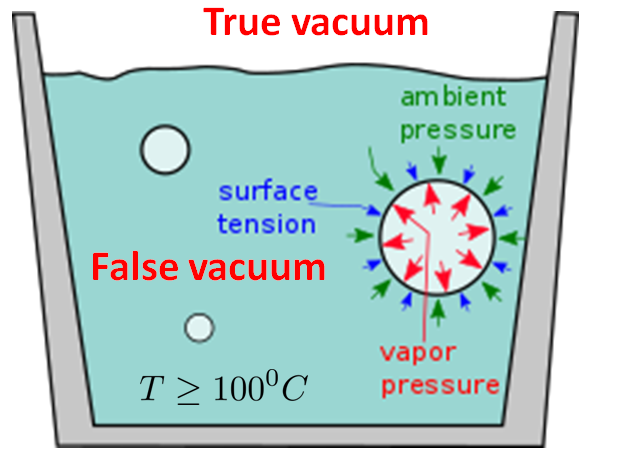}
		\caption{Nucleation of a vapor bubble in boiling water. If a thermal or quantum fluctuation creates a bubble that is too small, the inward force of surface tension dominates, and the bubble collapses. If it crosses a critical threshold, vapor pressure wins, and the bubble grows into the true vacuum state.}
	\end{figure}
	
	\subsection{The Higgs Field and the Early Universe}
	Similarly, cosmology heavily relies on these concepts when considering the Higgs field and the early universe. 
	The Higgs field $\phi(x)$ pervades all space and possesses a specific self-interacting potential $V(\phi)$. 
	Furthermore, the Higgs field $\phi(x)$ has charge under the weak force. 
	Since its vacuum expectation value is non-zero, the vacuum also has weak charge. 
	
	Let us conceptualize this using a generic potential for the scalar field. 
	The potential features a false vacuum and a true vacuum, separated by a potential barrier allowing for quantum tunneling. 
	
	\begin{figure}[h]
		\centering
		\includegraphics[width=0.8\textwidth]{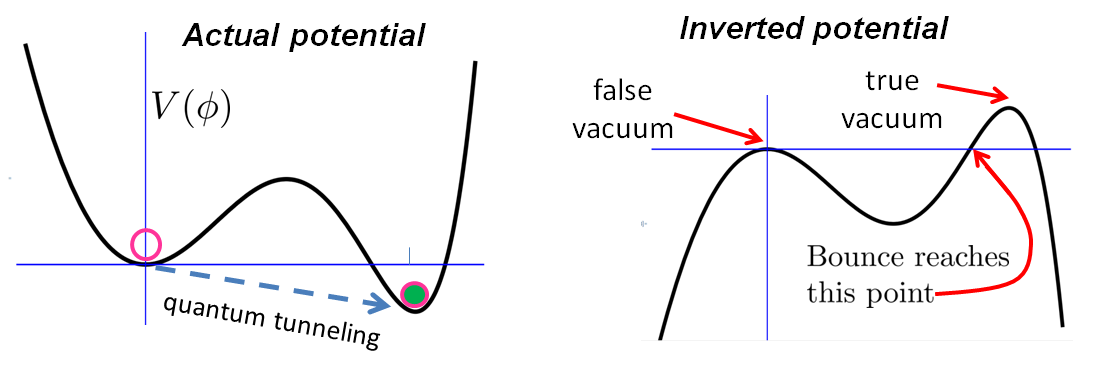}
		\caption{Left: The actual physical potential $V(\phi)$. Right: The inverted potential $-V(\phi)$ used in the Euclidean path integral, showing the classical bounce reaching its turning point.}
	\end{figure}
	
	\section{Relativistic QFT in Imaginary Time}
	We formulate the basics of relativistic QFT in imaginary time by applying a Wick rotation $t \to -i\tau$. 
	The Euclidean action is defined as $S_E[\phi] = \int d^4x_E \mathcal{L}_E$, where $d^4x_E = d\tau d^3x$. 
	The classical path, which extremizes the action, yields the equation of motion for the bounce path $\phi_b(\tau, \vec{x})$: 
	\begin{equation}
		\left( \frac{\partial^2}{\partial \tau^2} + \nabla^2 \right) \phi_b = V'(\phi_b)
	\end{equation}
	
	\section{The $O(4)$ Instanton and Bubble Nucleation}
	Because of the $\tau \to -\tau$ symmetry and spatial isotropy, we anticipate the $O(4)$ instanton, which depends only on the 4D radius $\rho = \sqrt{\tau^2 + x^2 + y^2 + z^2}$. 
	The PDE simplifies dramatically to a 1D ODE: 
	\begin{equation}
		\frac{d^2\phi_b}{d\rho^2} + \frac{3}{\rho} \frac{d\phi_b}{d\rho} = V'(\phi_b)
	\end{equation}
	
	Interpreting this radially, the term $\frac{3}{\rho} \frac{d\phi_b}{d\rho}$ acts as a fluid friction inversely proportional to "time" $\rho$. 
	The field starts at rest near the true vacuum (due to infinite initial friction), rolls rapidly across the barrier (forming the thin wall of the bubble), and comes to rest at the false vacuum as $\rho \to \infty$. 
	
	\begin{figure}[h]
		\centering
		\includegraphics[width=0.6\textwidth]{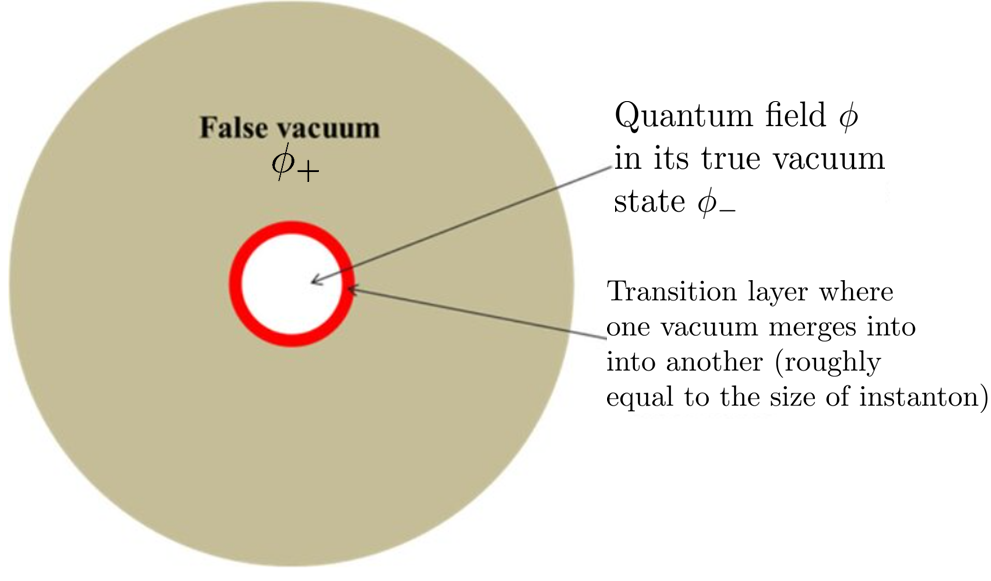}
		\caption{Nucleation of a true vacuum bubble within a false vacuum background. Inside the bubble, the field $\phi$ rests in its true vacuum state $\phi_-$.}
	\end{figure}
	
	\section{Decay Rate and The Thin-Wall Approximation}
	If the energy difference $\epsilon$ between the true and false vacuum is small, the $O(4)$ bounce resembles a sharp 4D spherical bubble of very large radius $R$. 
	We decompose the action $B(R)$ into volume and surface terms: 
	\begin{equation}
		B(R) = -\frac{1}{2}\pi^2 R^4 \epsilon + 2\pi^2 R^3 S_1
	\end{equation}
	Extremizing the action ($dB/dR = 0$) yields the critical radius $R_c = 3S_1 / \epsilon$. 
	Substituting this back gives the famous thin-wall result: 
	\begin{equation}
		B = \frac{27\pi^2 S_1^4}{2\epsilon^3}
	\end{equation}
	
	\section{The Expanding Bubble}
	Analytically continuing back to real Minkowski time ($t \to -i\tau$), the radial boundary $R^2 = r^2 - t^2$ describes the physical trajectory of the bubble wall: 
	\begin{equation}
		v(t) = \frac{dr}{dt} = \frac{t}{\sqrt{R^2 + t^2}} \xrightarrow{t \to \infty} 1
	\end{equation}
	The wall uniformly accelerates outward, asymptotically approaching the speed of light, driven by the pressure difference between the vacua. 
	
	\section{Gravitational Effects: Coleman-De Luccia Tunnelling}
	The inclusion of gravity minimally coupled to the field yields the Euclidean Einstein-Hilbert action. 
	Assuming an $O(4)$ invariant metric $ds^2 = d\rho^2 + a^2(\rho) d\Omega_3^2$, the coupled field equations for the bounce and the scale factor $a(\rho)$ become: 
	\begin{align}
		\phi_b'' + \frac{3a'}{a} \phi_b' &= V'(\phi_b) \\
		(a')^2 &= 1 + \frac{1}{3}\kappa a^2 \left( \frac{1}{2}(\phi_b')^2 - V(\phi_b) \right)
	\end{align}
	Once the false to true transition happens everything will be over, so bye-bye!

	\newpage
	\part{Instantons in Yang-Mills Gauge Theories}
	
	\section{Definition of the Yang-Mills Instanton}
	We now turn our attention to non-Abelian gauge theories. 
	The central concept governing this transition is formally defined as follows: 
	\begin{tcolorbox}[colback=blue!5,colframe=blue!40!black,title=Definition]
		\Large An instanton is a configuration of a Yang-Mills field that solves the equation of motion and has finite action. 
	\end{tcolorbox}
	
	\section{Review of Differential Forms}
	Recall that $U = U(x)$ is a transformation matrix with $U^\dagger(x)U(x) = 1$, and the covariant derivative is given by $D_\mu\phi(x) = \partial_\mu\phi(x) - iA_\mu(x)\phi(x)$. 
	For the kinetic term to transform covariantly, the gauge field $A_\mu$ must transform as: 
	\begin{equation}
		A_\mu \to U A_\mu U^\dagger + i U \partial_\mu U^\dagger
	\end{equation}
	
	\begin{figure}[h]
		\centering
		\includegraphics[width=\textwidth]{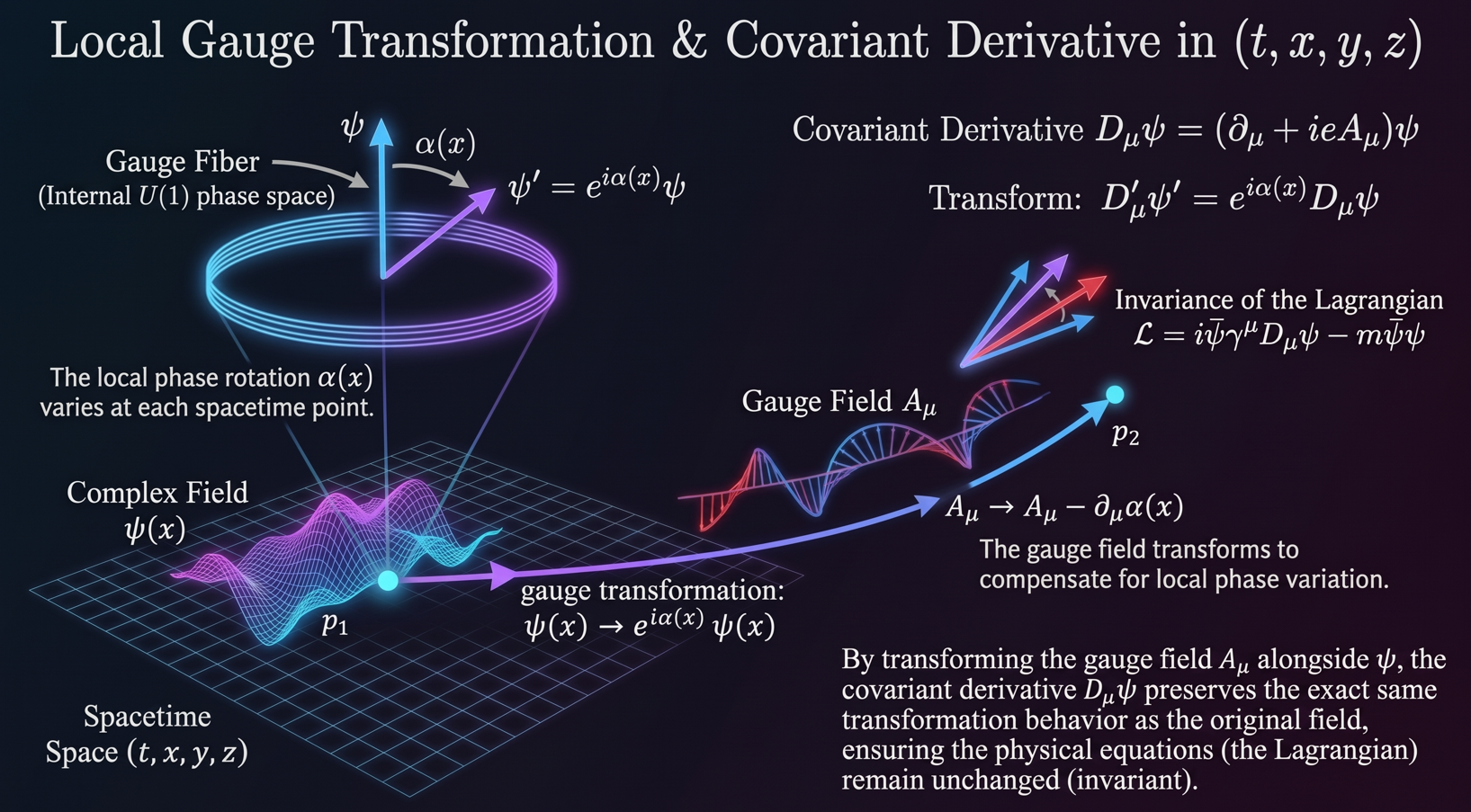}
		\caption{Visualizing a local gauge transformation. The gauge field $A_\mu$ acts as a connection along spacetime paths, transforming to compensate for local phase variations and preserving the invariance of the Lagrangian.}
	\end{figure}
	
	In the language of differential forms, we define the one-form $A \equiv A_\mu dx^\mu$ (hiding the $i$ by putting $-iA_\mu \equiv A_\mu$). 
	The transformation becomes $A \to U A U^\dagger + U dU^\dagger$. 
	
	\section{Constructing the Two-Form F}
	We define the two-form $F \equiv dA + AA = dA + A \wedge A$. 
	Under the gauge transformation, the inhomogeneous terms cancel perfectly: 
	\begin{equation}
		F \to U dA U^\dagger + U A A U^\dagger = U(dA + AA)U^\dagger = U F U^\dagger
	\end{equation}
	
	\section{Some Properties of F}
	We can recover space-time indices: 
	\begin{equation}
		F = \frac{1}{2}(\partial_\mu A_\nu - \partial_\nu A_\mu + [A_\mu, A_\nu]) dx^\mu \wedge dx^\nu \implies F_{\mu\nu} = \partial_\mu A_\nu - \partial_\nu A_\mu + [A_\mu, A_\nu]
	\end{equation}
	Restoring group indices using $A_\mu = A_\mu^a t^a$, we obtain $F_{\mu\nu}^a = \partial_\mu A_\nu^a - \partial_\nu A_\mu^a + f^{abc} A_\mu^b A_\nu^c$. 
	In 4-D, we can construct the topological invariant $\mathrm{tr}\, F^2$. 
	
	\section{The Euclidean Action}
	Recall we sent $t \to -i\tau$ so the kinetic and potential terms in the action $S$ carry the same sign: 
	\begin{equation}
		S[A] = \int d^4x \frac{1}{2g^2} \mathrm{tr}\, F^2
	\end{equation}
	To keep $S$ finite as $r \to \infty$, we must have $F \to 0$, which implies $A \to g dg^\dagger$ (a ``pure gauge'' configuration). 
	
	\section{Chern-Simons Form and Gauss's Theorem}
	The invariant $\mathrm{tr}\, F^2$ can be rewritten as an exact differential: $\mathrm{tr}\, F^2 = d \mathrm{tr}\left(A dA + \frac{2}{3} A^3\right)$. 
	Using Gauss's theorem and the pure gauge boundary condition $F=0$ on $S^3$, we find: 
	\begin{equation}
		\int_{\mathbb{R}^4} \mathrm{tr}\, F^2 = -\frac{1}{3} \mathrm{tr} \int_{S^3} (g dg^\dagger)^3
	\end{equation}
	
	\section{Multiple Vacua in SU(2)}
	The group manifold of $SU(2)$ is structurally equivalent to the 3-sphere $S^3$. 
	Different mappings $g^n$ from the spatial boundary onto the gauge group are valid because $\det g^n = 1$. 
	These different $n$'s characterize distinct topological mappings (winding numbers), corresponding to distinct vacuum states of the same minimal energy. 
	
	\begin{figure}[h]
		\centering
		\includegraphics[width=\textwidth]{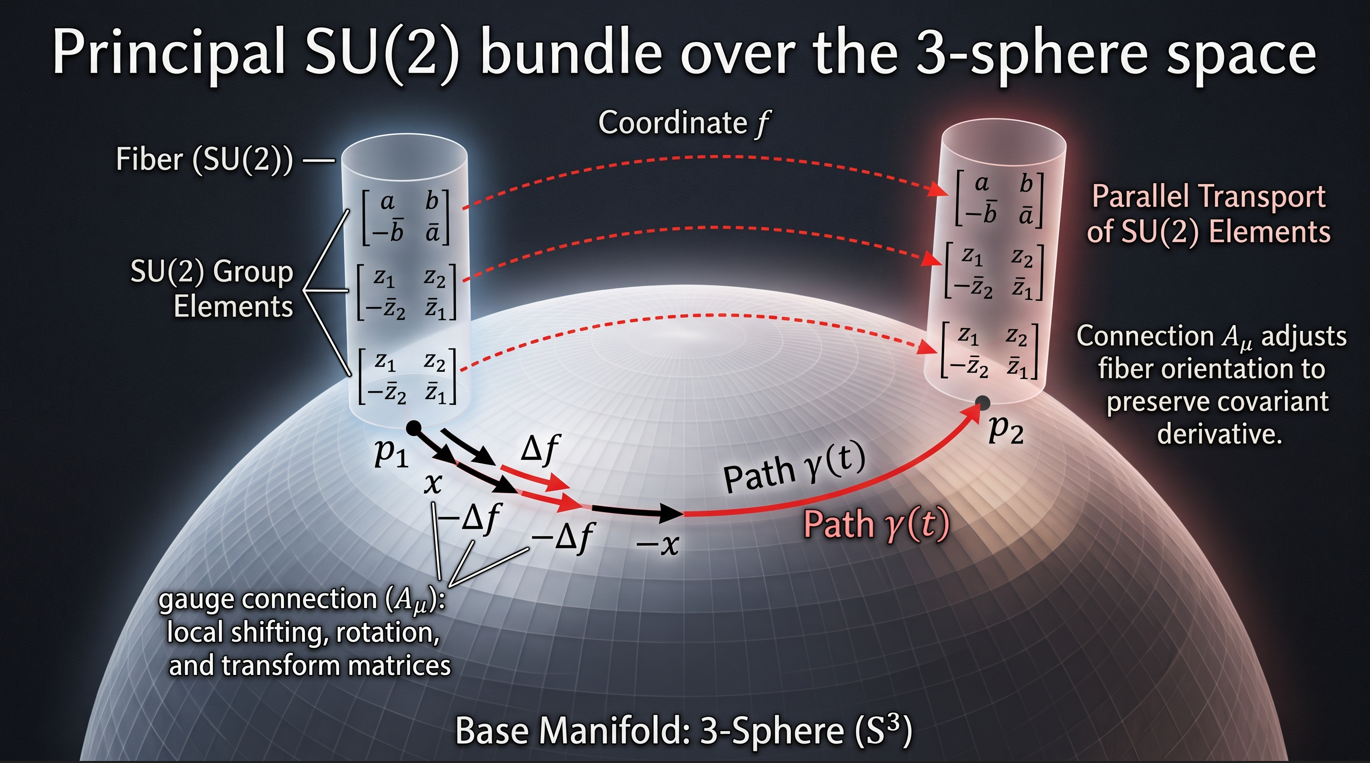}
		\caption{The principal $SU(2)$ fiber bundle over the 3-sphere spatial boundary. The connection $A_\mu$ governs the parallel transport of the internal $SU(2)$ elements.}
	\end{figure}
	
	\begin{figure}[h]
		\centering
		\begin{tikzpicture}[>=Stealth, scale=0.95]
			\fill[black!95, rounded corners=4pt] (-3.5, -3.0) rectangle (12.5, 5.8);
			\draw[step=1cm, white, opacity=0.015, ultra thin] (-3.5,-3.0) grid (12.5,5.8);
			
			\fill[cyan!10, opacity=0.05] (-2.5, -1.5) rectangle (10.5, 3);
			\draw[ultra thick, cyan!70!white, drop shadow={shadow xshift=0ex, shadow yshift=-1ex, shadow scale=1, opacity=0.3, fill=cyan}] 
			plot[domain=-2.5:10.5, samples=150] (\x, {1.5 - 1.5*cos(\x * 90)});
			
			\fill[cyan!30, opacity=0.15] 
			plot[domain=-2.5:10.5, samples=150] (\x, {1.5 - 1.5*cos(\x * 90)})
			-- (10.5, -2.5) -- (-2.5, -2.5) -- cycle;
			
			\coordinate (V1) at (0, 0); \coordinate (V2) at (4, 0); \coordinate (V3) at (8, 0);
			
			\filldraw[cyan!10, draw=cyan!80!white, ultra thick, drop shadow] (V1) circle (6pt) node[below=14pt, text=white, font=\Large] {$|n-1\rangle$};
			\filldraw[cyan!10, draw=cyan!80!white, ultra thick, drop shadow] (V2) circle (6pt) node[below=14pt, text=white, font=\Large] {$|n\rangle$};
			\filldraw[cyan!10, draw=cyan!80!white, ultra thick, drop shadow] (V3) circle (6pt) node[below=14pt, text=white, font=\Large] {$|n+1\rangle$};
			
			\draw[->, ultra thick, magenta!80!white, dashed] (V1) to[bend left=55] node[above=10pt, fill=black!95, inner sep=4pt, rounded corners, text=magenta!20!white, font=\Large] {Instanton ($Q = 1$)} (V2);
			\draw[->, ultra thick, magenta!80!white, dashed] (V2) to[bend left=55] node[above=10pt, fill=black!95, inner sep=4pt, rounded corners, text=magenta!20!white, font=\Large] {Tunneling Amplitude $\sim e^{-S_E}$} (V3);
			\draw[->, thick, green!60!black, dashed] (V2) to[bend left=40] node[below=8pt, fill=black!95, inner sep=3pt, rounded corners, text=green!50!white, font=\large] {Anti-instanton ($Q = -1$)} (V1);
			
			\node[text=white, font=\huge] at (4, 4.8) {$S_{\text{inst}} = \frac{8\pi^2}{g^2}$};
			\node[text=yellow!90!white, font=\LARGE] at (8.5, 4.8) {$|\theta\rangle = \sum_{n} e^{in\theta} |n\rangle$};
			\node[text=cyan!20!white, font=\Large] at (-0.5, 3.8) {$\displaystyle Q = \frac{g^2}{32\pi^2} \int d^4x \, F_{\mu\nu}^a \tilde{F}^{a\mu\nu}$};
			\node[text=cyan!50!white, font=\Large] at (4, -1.8) {Classical Vacuum: $F_{\mu\nu} = 0 \implies A_\mu = \frac{i}{g} U \partial_\mu U^{-1}$};
			\draw[->, thick, white!50] (8.4, 0) -- (10.0, 0) node[right=2pt, text=white!70, font=\large] {Chern-Simons $n$};
		\end{tikzpicture}
		\caption{The periodic vacuum structure of Yang-Mills theory. Distinct classical vacuum states are labeled by their winding number $n$. Instantons mediate tunneling transitions between these topologically distinct sectors.}
	\end{figure}
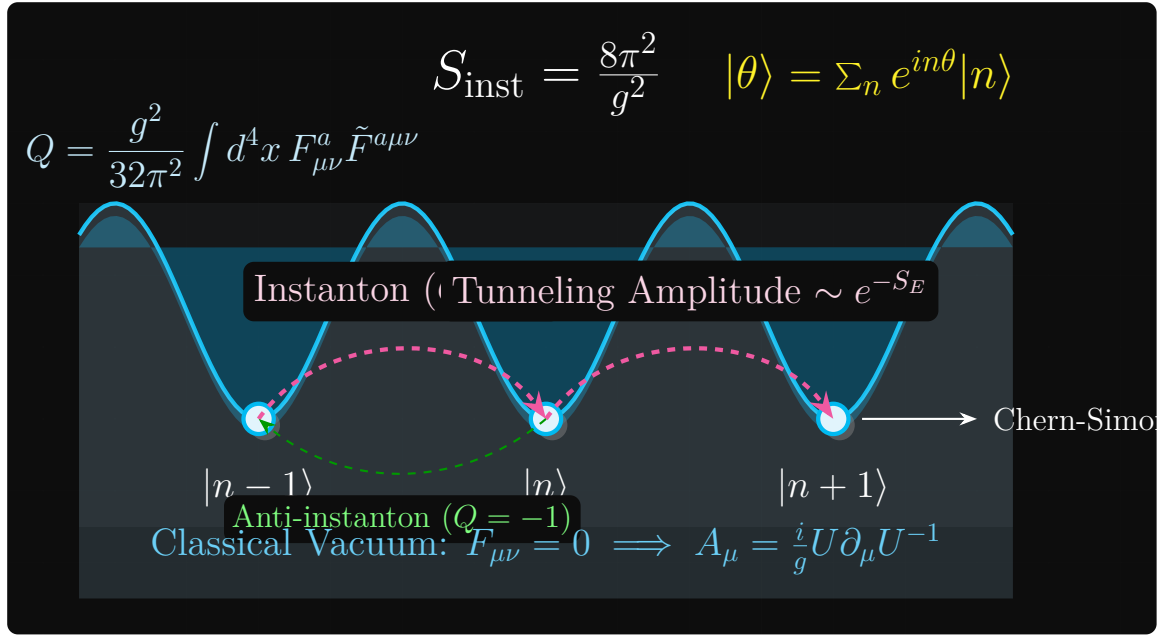
	
	\section{Application: The $U(1)_A$ Problem and the $\theta$-Vacuum}
	The theoretical formalism of instantons elegantly resolves one of the most glaring contradictions in early Quantum Chromodynamics (QCD): the $U(1)_A$ problem. 
	If the classical $U(1)_A$ axial symmetry of massless quarks were spontaneously broken by the QCD vacuum, Goldstone's theorem dictates the existence of a light pseudoscalar meson. 
	The candidate particle, the $\eta^\prime$ meson, possesses an anomalously heavy mass of roughly $958\text{ MeV}$, constituting a major crisis for the theory. 
	
	The classical $U(1)_A$ symmetry is anomalously broken at the quantum level. 
	The change in the chiral charge $\Delta Q_5$ is intimately tied to the topological winding number of the non-Abelian gauge field: 
	\begin{equation}
		\Delta Q_5 = 2 N_f \int d^4x \frac{g^2}{32\pi^2} F_{\mu\nu}^a \tilde{F}^{a \mu\nu} = 2 N_f \Delta n
	\end{equation}
	Because instantons mediate transitions between the $\vert{}n\rangle$ vacua ($\Delta n \neq 0$), the $U(1)_A$ symmetry is explicitly broken by these tunneling events. 
	
	\subsection{Topological Susceptibility and the Witten-Veneziano Relation}
	To make this mass generation mechanism quantitatively rigorous, we introduce the Witten-Veneziano relation, which defines the \textbf{topological susceptibility} ($\chi_t$): 
	\begin{equation}
		\chi_t = \int d^4x \, \langle q(x) q(0) \rangle
	\end{equation}
	where $q(x)$ is the topological charge density. 
	In the large-$N_c$ limit, the Witten-Veneziano formula establishes that the mass of the flavor-singlet $\eta^\prime$ is directly proportional to this gluonic susceptibility: 
	\begin{equation}
		m_{\eta^\prime}^2 = \frac{2 N_f}{f_\pi^2} \chi_t
	\end{equation}
	Through the anomalous coupling, the intense topological energy fluctuations of the Yang-Mills vacuum are injected directly into the flavor-singlet quark current. 
	This topological mass generation elevates the $\eta^\prime$ entirely out of the massless Goldstone multiplet, fully resolving the $U(1)_A$ problem. 
	
	Because tunneling connects all $\vert{}n\rangle$ states, the true ground state of the theory must be a gauge-invariant coherent superposition of all winding states, known as the $\theta$-vacuum. 
	This introduces a new, CP-violating topological term into the effective QCD Lagrangian: 
	\begin{equation}
		\mathcal{L}_\theta = \theta \frac{g^2}{32\pi^2} F_{\mu\nu}^a \tilde{F}^{a \mu\nu}
	\end{equation}
	Experimental bounds stringently demand that $\theta < 10^{-10}$. 
	Why this fundamental topological parameter is so vanishingly small is known as the Strong CP problem.

	\newpage
	\addcontentsline{toc}{part}{Recommended Readings}
	\renewcommand{\refname}{Recommended Readings}

\end{document}